\documentclass{aastex631}
\usepackage{subfigure}
\usepackage{graphicx}
\usepackage{multirow}
\usepackage{epstopdf}
\usepackage{float}
\usepackage[normalem]{ulem}
\usepackage{array}
\usepackage{threeparttable}
\usepackage{footnote}
\usepackage[figuresright]{rotating}
\usepackage{booktabs}
\usepackage{longtable}
\usepackage{threeparttable}
\usepackage[capitalize]{cleveref}

\useunder{\uline}{\ul}{}

\shorttitle{The properties of small magnetic flux ropes}
\shortauthors{Zhai et al.}

\graphicspath{{./}{figures/}}
\begin{document}

\title{The properties of small magnetic flux ropes inside the solar wind come from coronal holes, active regions, and quiet Sun}

\correspondingauthor{Hui Fu}
\email{fuhui@sdu.edu.cn}

\author{Changhao Zhai}
\affiliation{Shandong Key Laboratory of Optical Astronomy and Solar-Terrestrial Environment, Institute of Space Sciences, Shandong University, Weihai, Shandong, 264209, China}

\author[0000-0002-8827-9311]{Hui Fu}
\affiliation{Shandong Key Laboratory of Optical Astronomy and Solar-Terrestrial Environment, Institute of Space Sciences, Shandong University, Weihai, Shandong, 264209, China}

\author{Jiachen Si}
\affiliation{Shandong Key Laboratory of Optical Astronomy and Solar-Terrestrial Environment, Institute of Space Sciences, Shandong University, Weihai, Shandong, 264209, China}

\author[0000-0002-2358-5377]{Zhenghua Huang}
\affiliation{Shandong Key Laboratory of Optical Astronomy and Solar-Terrestrial Environment, Institute of Space Sciences, Shandong University, Weihai, Shandong, 264209, China}

\author[0000-0001-8938-1038]{Lidong Xia}
\affiliation{Shandong Key Laboratory of Optical Astronomy and Solar-Terrestrial Environment, Institute of Space Sciences, Shandong University, Weihai, Shandong, 264209, China}

\begin{abstract}
The origination and generation mechanisms of small magnetic flux ropes (SFRs), which are important structures in solar wind, 
are not clearly known. In present study, 1993 SFRs immersed in coronal holes, active regions, and quiet Sun solar wind are analyzed and compared. We find that the properties of SFRs immersed in three types of solar wind are significantly different. The SFRs are further classified into hot-SFRs, cold-SFRs, and normal-SFRs, according to whether the O$^{7+}$/O$^{6+}$ is {30$\%$} elevated or dropped inside SFRs {as compared with background solar wind}. Our studies show that the parameters of normal-SFRs are similar to background in all three types of solar wind. 
The properties of hot-SFRs and cold-SFRs seem to be lying in two extremes.
Statistically, the hot-SFRs (cold-SFRs) are associated with longer (shorter) duration, lower (higher) speeds and proton temperatures, higher (lower) charge states, helium abundance, and FIP bias as compared with normal-SFRs and background solar wind.
The anti-correlations between speed and O$^{7+}$/O$^{6+}$ inside hot-SFRs (normal-SFRs) are different from (similar to) those in background solar wind. 
%The properties of hot-SFRs can be explained reasonably by the notions that they come from the streamers associated with plasma blobs and/or are produced by small-scale activities on the Sun. The cold-SFRs should also originate from the Sun and be associated with lower plasma temperature in source regions. Both hot-SFRs and cold-SFRs could also be formed by magnetic erosions of ICMEs. 
{Most of hot-SFRs and cold-SFRs should come from the Sun. Hot-SFRs may come from streamers associated with plasma blobs and/or small-scale activities on the Sun. Cold-SFRs may be accompanied by small-scale eruptions with lower-temperature materials. Both hot-SFRs and cold-SFRs could also be formed by magnetic erosions of ICMEs that do not contain or contain cold-filament materials.}
%The properties of hot-SFRs \textbf{could} be explained reasonably by the notions that they come from the streamers associated with plasma blobs and/or be produced by small-scale activities on the Sun. The cold-SFRs \textbf{may} originate from the Sun and be associated with lower \textbf{electron} temperature in source regions. \textbf{Both hot-SFRs and cold-SFRs could also be formed by magnetic erosions of ICMEs.}
The characteristics of normal-SFRs can be explained reasonably by the two originations, from the Sun and generated in the heliosphere both.
%The normal-SFRs may come from the Sun and be generated in the heliosphere both.
% 
%in different types of solar wind 
%\textbf{Most of hot-SFRs and cold-SFRs come from the Sun. The hot-SFRs may come from the streamers associated with plasma blobs and/or be produced by small-scale activities on the Sun. The cold-SFRs may also originate from the Sun and be associated with lower plasma temperature in source regions. Both hot-SFRs and cold-SFRs could also be formed by magnetic erosions of ICMEs that do not contain or contain cold-filament or prominence materials. The normal-SFRs may come from the Sun and be generated in the heliosphere both.}

\end{abstract}


\section{Introduction}\label{sec:intro}
Magnetic flux ropes, which are important structures in the solar wind (SW), can be detected at different heliocentric distances \citep{2010JGRA..115.8102C, 2020ApJ...894...25C,2022ApJ...924...43C,2021A&A...650A..12Z}. It is a twisted magnetic field structure that rotates along an axis. The duration of magnetic flux ropes spreads over a wide range, from several minutes to dozens of hours. 
Generally, magnetic flux ropes are categorized into large flux ropes and small flux ropes (SFRs) according to duration or diameter of the flux ropes.
Large flux ropes, also known as magnetic clouds (MCs), are generally associated with interplanetary coronal mass ejections (ICMEs). The radial scale of MCs ranges from 0.2 to 0.4 AU, with duration lasting for hours to dozens of hours \citep{1990JGR....9511957L, 2000GeoRL..27...57M}. The properties, origination, and generation mechanisms of MCs have been comprehensively analyzed. Generally, MCs are associated with stronger magnetic field strengths, lower proton temperatures, and lower plasma beta in comparison with solar wind \citep{1990JGR....9511957L, 1991pihp.book....1B, 2003JGRA..108.1156C}. Statistically, the charge states, first ionization potential (FIP) bias, and helium abundance (A$_{He}$) inside MCs are higher than those inside the background solar wind \citep{2004ApJ...612.1171W, 2010JGRA..115.4103R, 2017SSRv..212.1159M, 2018SoPh..293..122O, 2021SoPh..296..111S, 2022ApJ...940..103S}. 
There is a consensus that MCs originate from the Sun and are closely related to the violent eruption events on the Sun. SFRs typically last no more than 12 hours and have diameters less than 0.2 AU \citep{2008JGRA..11312105F, 2015ApJ...809..112F, 2013AIPC.1539..311Y,2016JGRA..121.5005Y,2017JGRA..122.6927H}. The origination and generation mechanisms of SFRs are still not clear.

In general, it is believed that SFRs may originate from the Sun and/or be generated in the heliosphere \citep{1995JGR...10019903M, 2000GeoRL..27...57M, 2008JGRA..113.9105C, 2009SoPh..256..307R, 2010JGRA..115.4103R,2010JGRA..115.4104R,2011ApJ...734....7R, 2015ApJ...809..112F, 2016JGRA..121.5005Y, 2017ApJ...835L...7S, 2019ApJ...882...51S, 2018ApJS..239...12H, 2018JGRA..123.7167H, 2018ApJ...852L..23Z, 2020ApJ...904..122X, 2020ApJ...899L..29H, 2021SoPh..296..148C}. 
Some studies indicate that the SFRs are generated in the heliosphere during the propagation of solar wind.
As part of analyses demonstrate that the in-situ properties of SFRs and MCs are different.
A small magnetic flux rope near the heliospheric current sheet (HCS) with a radial scale of about 0.05 AU was first detected by Ulysses at 5 AU \citep{1995JGR...10019903M}. The properties of this SFR are different from those of MCs. The SFR is associated with high densities and plasma beta, and high proton temperatures. Hence, they implied that the SFR is generated by multiple magnetic reconnections near the HCS in interplanetary space.
The six SFRs observed by the Wind and IMP 8 spacecraft were analyzed by \citet{2000GeoRL..27...57M}. They suggested that the SFRs are generated by magnetic reconnection in the solar wind rather than originating from the solar corona, as the expansion signatures and proton temperature of SFRs are not the same as MCs.
The duration distributions, wall-to-wall time distributions, and axial directions of SFRs are also analyzed.
\citet{2008JGRA..113.9105C} identified 68 flux ropes with a duration ranging from 39 minutes to 12.2 hours. The statistical results showed that the duration of flux ropes presents a bimodal distribution. The majority of SFR events have a duration of less than 4 hours. In contrast, the duration distribution of MCs peaks at 12 to 16 hours. Thus, they suggested that the sources of SFRs and MCs are not identical and the origin of SFRs may be associated with reconnection in the solar wind.
The wall-to-wall time of SFRs displays a power-law distribution and the axial current density of SFRs presents a non-Gaussian distribution \citep{2018ApJ...852L..23Z}. Hence, they suggested that the SFRs are generated in the heliosphere and are closely connected with the local MHD turbulence.
The relationship between the axial directions of SFRs and the directions of HCSs was analyzed by \citet{2021FrP.....9..637L}. They found that the axial directions of SFRs are generally parallel with HCSs. Therefore, they suggested that the SFRs are generated in the heliosphere near HCSs.

Other researches support the idea that the SFRs originate from the Sun. 
The in-situ and remote joint observations provide direct evidence for the notion that SFRs can come from the Sun. The Solar TErrestrial RElations Observatory \citep[STEREO;][]{2008SSRv..136....5K} which includes two nearly identical spacecraft (A and B) {provide} stereoscopic observations of activities on the Sun and in the heliosphere. An SFR with a radial scale of about 0.08 AU was detected by STEREO-A before a corotating interaction region (CIR). The SFR and CIR were imaged by the Heliospheric Imager (HI) instrument onboard STEREO-B from the Sun to STEREO-A \citep{2009SoPh..256..307R}. This demonstrated that the SFR should originate from the Sun. The remote observations taken by HI instruments onboard the STEREO spacecraft found that intermittent outflows transport along the extension of helmet streamers. Some twisted structures which may be SFRs are observed near streamers in the remote images \citep{2010JGRA..115.4103R}. Further, \citet{2010JGRA..115.4104R} compared the remote observations taken by HI instruments with in situ measurements detected by Advanced Composition Explorer %\footnote{\url{http://www.srl.caltech.edu/ACE/}} 
\citep[ACE;][]{1998SSRv...86....1S} and STEREO. The above joint observations confirmed the association between the transits and SFRs. In addition, the SFRs detected in situ by STEREO and ACE were traced back by the coronagraph images. \citet{2011ApJ...734....7R} found that the SFRs are associated with either small or large mass ejecta on the Sun. The small {ejecta} generally appear at the tip of streamers. Further observations suggested that at least part of SFRs come from the Sun and should be generated by magnetic reconnection at the tip of coronal helmet streamers \citep{2017ApJ...835L...7S,2019ApJ...882...51S}. 

Some comparisons between SFRs and MCs suggest that SFRs may come from the Sun.
\citet{2007JGRA..112.2102F} identified 144 magnetic flux ropes with a radial scale of about 0.004-0.6 AU (the duration from minutes to dozens of hours) from the measurements of Wind. The size and energy distributions of flux ropes are continuous. There is no doubt that the large-scale flux ropes (MCs) come from the Sun. Therefore, they suggested that the SFRs also originate from the Sun and are produced by small-scale eruptions.
The 125 small and intermediate magnetic flux ropes with duration ranging from 0.45 to 10.87 hours detected by Wind were analyzed by \citet{2008JGRA..11312105F}. They found that the average magnetic field strengths of SFRs are larger than those of background solar wind, and the axial orientations of SFRs are almost the same as those of MCs. The above statistical results indicated that the SFRs also originate from the Sun. 
\citet{2015ApJ...809..112F} identified 24 SFRs from the measures of the ACE with a duration from 4 to 12 hours. The statistical results showed that the SFRs are associated with the same characteristics of high charge states and helium abundance as MCs. Hence, they suggested that the SFRs and MCs should be produced by the same process on the Sun.
Comparing the temperature, density, helium abundance, and charge states inside heliospheric plasma sheet (HPS) and an SFR near the HPS, \citet{2017JGRA..122.6927H} found that the properties of HPS and the SFR near HPS are similar. They implied that both SFR and HPS originate from the corona and the SFR is generated by interchange reconnection in the streamer belt.

Many studies demonstrate that the SFRs can be generated both on the Sun and in the heliosphere.
The statistical results indicated that the SFRs may originate from both the Sun and the heliosphere, as the charge states are higher inside part of SFRs \citep{2016JGRA..121.5005Y,2018JGRA..123.7167H}. \citet{2016JGRA..121.5005Y} found that only about 5$\%$ of SFRs present enhanced iron average charge states ($Q_{Fe}$) using the STEREO measurements. The distributions of $Q_{Fe}$ are also investigated based on 25 SFRs detected by ACE \citep{2018JGRA..123.7167H}. They found that abZout one-third (8/25) of SFRs associated with higher $Q_{Fe}$. 
The properties of SFRs near and far from HCSs, with and without counterstreaming suprathermal electrons (CSEs) are also analyzed.
\citet{2015JGRA..12010175F} classified the SFRs into two types based on the distance between the SFRs and HCSs. They found that the SFRs far from HCSs are generally (50 of 57 cases) associated with CSEs. In contrast, only 60$\%$ of SFRs near HCSs are accompanied by CSEs. They suggested that the SFRs might be formed in the corona and near HCSs in interplanetary space both. 
\citet{2020ApJ...899L..29H} classified SFRs into two types, with CSEs and far from HCSs, and without CSEs and near HCSs. They found that the properties of the two types of SFRs are different. The parameters of SFRs with CSEs and far from HCSs are almost the same as those of MCs. The SFRs with CSEs and far from HCSs are generally associated with higher FIP bias effect and charge states as compared with SFRs without CSEs and near HCSs. Accordingly, they suggested that the SFRs can be generated both on the Sun and in the heliosphere.
\citet{2021SoPh..296..148C} analyzed the distribution characteristics of suprathermal electrons inside SFRs. They found that the bidirectional beams of suprathermal electrons inside SFRs are presented in only 10.7$\%$ of cases. The above statistical results demonstrated that the properties of suprathermal electrons inside the majority of SFRs are not the same as MCs.
The SFRs inside ICMEs, solar wind, and stream interaction regions (SIRs) are statistically analyzed by \citet{2020ApJ...904..122X}. They found that the properties of SFRs inside ICMEs are significantly different from the SFRs inside the solar wind and SIRs. The probabilities of SFRs with strong magnetic fields, significant expansion, high iron charge states, and CSEs are higher in ICMEs than those in solar wind and SIRs. Therefore, they suggested that the origination mechanisms for the majority of SFRs are not the same as that of MCs, and only a small part of SFRs come from the Sun.
In addition, SFRs can also be produced by the magnetic erosion processes. Previous works demonstrated that magnetic reconnection can occur at the front and/or at the rear of MCs \citep{2006A&A...455..349D,2007SoPh..244..115D,2012JGRA..117.9101R,2014JGRA..119.7088J,2014JGRA..119...26L} and SFRs \citep{2009ApJ...705.1385F, 2010ApJ...720..454T}. {The magnetic erosion effect of MCs can reduce 40$\%$ of the total azimuthal magnetic flux \citep{2015JGRA..120...43R}.} The above processes can erode parts of magnetic field structure of MCs {to generate} SFRs. Therefore, magnetic erosion in the heliosphere should also {contribute} to the formation of SFRs. This means that the formation of SFRs should be associated with the processes both on the Sun and in the heliosphere.

Generally, the above results are based on a case study or a low number of cases. Considering that the monthly number of SFRs is about 300-500 (100) during solar maximum and decline (minimum) phases, based on one-point observation by Wind spacecraft at 1 AU (see Figure 3(a) in \citet{2018ApJS..239...12H} and Figure 1 in \citet{2018ApJ...852L..23Z}).
\citet{2018ApJS..239...12H} developed an automated detection method for SFRs based on the Grad-Shafranov (GS) reconstruction technique. They extracted 74241 SFRs in the solar wind measured by Wind during solar cycles 23 and 24, with a duration of 9 to 361 minutes. The statistical analysis using the above SFR list showed that the SFRs appear more frequently near the HCSs \citep{2018ApJS..239...12H}. 
It is worth noting that the occurrence of SFRs is generally positively correlated with solar activities \citep{2018ApJS..239...12H,2018ApJ...852L..23Z}. However, the above positive correlation does not only mean that the SFRs originate from the Sun, considering the local processes to produce the SFRs in the heliosphere can also be modulated by solar activity \citep{2018ApJS..239...12H,2018ApJ...852L..23Z}.
The automated detection method \citep{2018ApJS..239...12H} based on the Grad–Shafranov reconstruction technique \citep{2017SoPh..292..116H} is also applied to the data detected by Helios 1, 2 and Voyager spacecraft. The statistical results demonstrated that the event counts of SFRs decrease with increasing radial distance from 0.3 to about 7 AU in the ecliptic plane \citep{2020ApJ...894...25C}. 

The properties of SFRs inside solar wind from different source regions have not been analyzed and compared systematically. The three main solar wind source regions, coronal holes (CHs), active regions (ARs), and quiet Sun (QS), differ greatly in plasma properties, magnetic field structures, and magnetic activities. The CH (AR and QS) regions are dominated by open (closed) magnetic field structures \citep{2004SoPh..225..227W, 2010ApJ...719..131I, 2014A&ARv..22...78W}. The temperature of ARs is higher than that of CH regions, with the temperature of QS regions lying in between.
In addition, explosive events are more frequently occurring in the active regions as the magnetic field is stronger and the evolution is faster \citep{2017MNRAS.464.1753H}. 
Hence, the charge states, helium abundance, and FIP bias of solar wind originating from different source regions are significantly different \citep{2015SoPh..290.1399F, 2017ApJ...836..169F,2018MNRAS.478.1884F,2016SSRv..201...55A}. 
Therefore, it is crucial to analyze the properties of SFRs inside the solar wind originating from different source regions. This can promote the understanding of the origination and generation mechanisms of SFRs.
The SFRs detected by the ACE spacecraft are recognized by the automated detection method based on the Grad-Shafranov reconstruction technique \citep{2017SoPh..292..116H, 2017SoPh..292..171H, 2018ApJS..239...12H, 2019ApJ...881...58C}. This is a big progress in the study of SFRs as the number of recognized events is huge. The monthly number of recognized SFRs is about several (one) hundreds during solar maximum and decline (minimum) phases. This makes the statistical analysis on SFR properties, in particular for compositional signatures, possible.

In the present study, the properties of SFRs inside the solar wind originating from CH, AR, and QS regions are analyzed. The charge states, FIP bias, and helium abundance are mainly analyzed as they carry the information of source regions in the Sun \citep{2017SSRv..212.1159M}.
Firstly, the properties of SFRs immersed in CH, AR, and QS solar wind and their associated background are analyzed and compared. The statistical analysis shows that the compositional properties of the three types of SFRs are significantly different. There are only small differences between the compositional parameters inside SFRs and their associated background solar wind. 
Secondly, the SFRs are further classified into hot-SFRs, cold-SFRs, and normal-SFRs, according to whether or not the O$^{7+}$/O$^{6+}$ is clearly elevated or dropped inside SFRs {as} compared with associated background solar wind. The properties of three types SFRs immersed in CH, AR, and QS solar wind are analyzed and compared. The statistical results demonstrate that the properties of hot-SFRs and cold-SFRs are significantly different from the normal-SFRs. The properties of hot-SFRs and cold-SFRs {seem} to be lying in the two extremes compared with normal-SFRs and background solar wind. Finally, the distributions in the space of speed versus O$^{7+}$/O$^{6+}$ for hot-SFRs, cold-SFRs, normal-SFRs, and background solar wind are investigated. The distributions of hot-SFRs are different from those of cold-SFRs, normal-SFRs, and background solar wind. The relevant results will enhance our understanding of SFR properties and help us understand the origination and generation mechanisms of SFRs.
This paper is structured as follows. The SFR list, the data, and analysis methods are introduced in Section~\ref{sec:style}. In Section~\ref{sec:floats}, the results are presented and discussed. The statistical results are concluded in Section~\ref{sec:summary}.

\section{Data and analysis} \label{sec:style}

In the present study, the SFR ACE EVENT LIST presented in the SMALL-SCALE MAGNETIC FLUX ROPE DATABASE\footnote{\url{http://fluxrope.info/}} is used. The SFRs are recognized by an automated detection method based on the Grad-Shafranov reconstruction technique \citep{2017SoPh..292..116H,2017SoPh..292..171H,2018ApJS..239...12H,2019ApJ...881...58C}. The SFRs studied in this paper are detected by ACE from 1999 to 2008, including solar maximum (1999-2002), decline (2003-2006), and minimum (2007-2008) phases of solar cycle 23.
The SFRs ranging from 1 to 12 hours are chosen as the time resolution of compositional parameters is 1 hour.
The SFRs are categorized into three types based on the source region types of associated solar wind. \citet{2015SoPh..290.1399F} traced the solar wind near the Earth back to the Sun. The source regions were categorized into coronal holes, active regions, and quiet Sun regions based on the photospheric magnetic field structures and EUV images. Then the solar wind was classified into three groups based on the source region types. The statistical analysis demonstrated that the in-situ parameters of the three types of solar wind are significantly different \citep{2017ApJ...836..169F,2018MNRAS.478.1884F}. 

In total, 1993 SFRs are analyzed in the present study. The numbers of SFRs immersed in CH, AR, and QS solar wind are 652, 988, and 353, respectively. 
The background solar wind is defined as 12 hours preceding the leading edges and 12 hours behind the trailing edges of SFRs.
The intervals occupied by other SFRs in the defined background solar wind are also excluded. Then, the properties, especially the compositional signatures of SFRs and their associated background in the CH, AR, and QS solar wind are analyzed and compared. Finally, the distributions in the space of speed versus O$^{7+}$/O$^{6+}$ for the SFRs immersed in different types of solar wind are also investigated.

The in-situ parameters of SFRs and solar wind are measured by the ACE spacecraft.%\textbf{\footnote{\url{http://www.srl.caltech.edu/ACE/}}} 
The speed, proton number density, helium abundance, and proton temperature of solar wind are taken by Solar Wind Electron Proton Alpha Monitor \citep[SWEPAM;][]{1998SSRv...86..563M} and the magnetic field is detected by Magnetic Field Experiment \citep[MAG;][]{1998SSRv...86..613S}.
The charge states (O$^{7+}$/O$^{6+}$, C$^{6+}$/C$^{5+}$, and C$^{6+}$/C$^{4+}$) and FIP bias (Fe/O, Mg/O) are measured by Solar Wind Ion Composition Spectrometer \citep[SWICS;][]{1998SSRv...86..497G}. 
The SFRs longer than 1 hour are analyzed as the time resolution of charge states and FIP bias is 1 hour. 
If one hour (from t0 to t0+1) is totally covered by an SFR then the associated value of the hour is adopted. In general, the two edges of the SFRs do not exactly occupy the whole hour (from t0 to t0+1). The value of an hour (from t0 to t0+1) at starting or ending part of an SFR is adopted if more than half of the hour is occupied by the SFR.

\section{Results and Disscussion} \label{sec:floats}

\subsection{The properties of SFRs and their associated background inside CH, AR, and QS solar wind}

\begin{table}[]
\centering
\caption{The paramters of SFRs and background in CH, AR, and QS solar wind }
\label{tab:1}
\resizebox{\columnwidth}{!}{%
\begin{threeparttable} 
\begin{tabular}{cccccc}
\hline
 &
   &
  \textbf{MAX} &
  \textbf{DEC} &
  \textbf{MIN} &
  \textbf{TOTAL} \\ \hline
\multirow{4}{*}{\textbf{Number of SFRs}} &
  CH &
  207 &
  350 &
  95 &
  652 \\
 &
  AR &
  689 &
  257 &
  42 &
  988 \\
 &
  QS &
  160 &
  136 &
  57 &
  353 \\
 &
  ALL &
  1056 &
  743 &
  194 &
  1993 \\ \hline
\multirow{4}{*}{\textbf{Duration (hour) \tnote{a}}} &
  CH &
  1.89 $\pm$ 1.22 &
  1.85 $\pm$ 1.13 &
  1.82 $\pm$ 1.25 &
  1.86 $\pm$ 1.18 \\
 &
  AR &
  2.26 $\pm$ 1.65 &
  2.12 $\pm$ 1.46 &
  1.76 $\pm$ 1.19 &
  2.20 $\pm$ 1.59 \\
 &
  QS &
  2.09 $\pm$ 1.42 &
  2.01 $\pm$ 1.35 &
  2.26 $\pm$ 1.87 &
  2.09 $\pm$ 1.47 \\
 &
  ALL &
  2.16 $\pm$ 1.55 &
  1.97 $\pm$ 1.30 &
  1.94 $\pm$ 1.46 &
  2.07 $\pm$ 1.45 \\ \hline
\multirow{4}{*}{\textbf{O$^{7+}$/O$^{6+}$ \tnote{c}}} &
  CH (SW)\tnote{b} &
  0.19 $\pm$ 0.16 (0.15 $\pm$ 0.13) &
  0.14 $\pm$ 0.12 (0.11 $\pm$ 0.13) &
  0.08 $\pm$ 0.07 (0.06 $\pm$ 0.07) &
  0.15 $\pm$ 0.13 (0.12 $\pm$ 0.12) \\
 &
  AR (SW) &
  0.30 $\pm$ 0.20 (0.25 $\pm$   0.20) &
  0.22 $\pm$ 0.15 (0.19 $\pm$ 0.15) &
  0.11 $\pm$ 0.08 (0.09 $\pm$ 0.08) &
  0.27 $\pm$ 0.19 (0.22 $\pm$ 0.19) \\
 &
  QS (SW) &
  0.27 $\pm$ 0.26 (0.21 $\pm$   0.19) &
  0.19 $\pm$ 0.10 (0.16 $\pm$ 0.13) &
  0.04 $\pm$ 0.04 (0.06 $\pm$ 0.05) &
  0.20 $\pm$ 0.20 (0.16 $\pm$ 0.16) \\
 &
  ALL (SW) &
  0.27 $\pm$ 0.21 (0.22 $\pm$   0.19) &
  0.18 $\pm$ 0.13 (0.15 $\pm$ 0.14) &
  0.07 $\pm$ 0.07 (0.06 $\pm$ 0.07) &
  0.22 $\pm$ 0.19 (0.18 $\pm$ 0.17) \\ \hline
\multirow{4}{*}{\textbf{C$^{6+}$/C$^{5+}$ \tnote{c}}} &
  CH (SW) &
  0.86 $\pm$ 0.54 (0.80 $\pm$   0.54) &
  0.75 $\pm$ 0.53 (0.63 $\pm$ 0.54) &
  0.56 $\pm$ 0.39 (0.43 $\pm$ 0.42) &
  0.76 $\pm$ 0.52 (0.65 $\pm$ 0.54) \\
 &
  AR (SW) &
  1.19 $\pm$ 0.77 (1.11 $\pm$   0.76) &
  1.20 $\pm$ 0.74 (1.01 $\pm$ 0.67) &
  0.79 $\pm$ 0.49 (0.63 $\pm$ 0.53) &
  1.18 $\pm$ 0.75 (1.06 $\pm$ 0.73) \\
 &
  QS (SW) &
  1.11 $\pm$ 0.69 (1.08 $\pm$   0.81) &
  1.12 $\pm$ 0.54 (0.96 $\pm$ 0.61) &
  0.43 $\pm$ 0.30 (0.52 $\pm$ 0.40) &
  1.00 $\pm$ 0.64 (0.92 $\pm$ 0.70) \\
 &
  ALL (SW) &
  1.12 $\pm$ 0.73 (1.04 $\pm$   0.74) &
  0.99 $\pm$ 0.65 (0.82 $\pm$ 0.63) &
  0.56 $\pm$ 0.41 (0.50 $\pm$ 0.45) &
  1.02 $\pm$ 0.70 (0.89 $\pm$ 0.69) \\ \hline
\multirow{4}{*}{\textbf{C$^{6+}$/C$^{4+}$ \tnote{c}}} &
  CH (SW) &
  3.02 $\pm$ 2.55 (2.75 $\pm$   2.41) &
  2.53 $\pm$ 2.36 (2.09 $\pm$ 2.42) &
  1.35 $\pm$ 1.51 (1.01 $\pm$ 1.34) &
  2.52 $\pm$ 2.38 (2.13 $\pm$ 2.35) \\
 &
  AR (SW) &
  4.48 $\pm$ 3.86 (4.04 $\pm$   3.85) &
  4.41 $\pm$ 4.45 (3.61 $\pm$ 3.96) &
  2.08 $\pm$ 1.79 (1.71 $\pm$ 1.82) &
  4.38 $\pm$ 3.99 (3.80 $\pm$ 3.84) \\
 &
  QS (SW) &
  4.12 $\pm$ 3.29 (3.74 $\pm$   3.04) &
  3.89 $\pm$ 2.50 (3.21 $\pm$ 2.85) &
  0.85 $\pm$ 0.83 (1.16 $\pm$ 1.40) &
  3.46 $\pm$ 2.97 (3.01 $\pm$ 2.88) \\
 &
  ALL (SW) &
  4.18 $\pm$ 3.63 (3.73 $\pm$   3.52) &
  3.48 $\pm$ 3.42 (2.79 $\pm$ 3.23) &
  1.32 $\pm$ 1.45 (1.21 $\pm$ 1.51) &
  3.67 $\pm$ 3.51 (3.07 $\pm$ 3.33) \\ \hline
\multirow{4}{*}{\textbf{FIP bias (Fe/O) \tnote{d}}} &
  CH (SW) &
  2.49 $\pm$ 1.15 (2.33 $\pm$   0.94) &
  2.14 $\pm$ 0.95 (1.99 $\pm$ 0.84) &
  2.07 $\pm$ 1.08 (1.85 $\pm$ 0.81) &
  2.25 $\pm$ 1.05 (2.08 $\pm$ 0.89) \\
 &
  AR (SW) &
  3.12 $\pm$ 1.63 (2.89 $\pm$   1.49) &
  2.76 $\pm$ 1.33 (2.53 $\pm$ 1.33) &
  2.22 $\pm$ 1.00 (2.09 $\pm$ 0.86) &
  3.00 $\pm$ 1.55 (2.75 $\pm$ 1.44) \\
 &
  QS (SW) &
  2.75 $\pm$ 1.45 (2.46 $\pm$   1.05) &
  2.36 $\pm$ 0.73 (2.16 $\pm$ 0.76) &
  1.83 $\pm$ 0.61 (1.75 $\pm$ 0.81) &
  2.44 $\pm$ 1.15 (2.20 $\pm$ 0.94) \\
 &
  ALL (SW) &
  2.96 $\pm$ 1.55 (2.71 $\pm$   1.36) &
  2.41 $\pm$ 1.11 (2.20 $\pm$ 1.05) &
  2.02 $\pm$ 0.94 (1.86 $\pm$ 0.84) &
  2.68 $\pm$ 1.40 (2.41 $\pm$ 1.23) \\ \hline
\multirow{4}{*}{\textbf{FIP bias (Mg/O) \tnote{e}}} &
  CH (SW) &
  1.81 $\pm$ 0.55 (1.75 $\pm$   0.48) &
  1.60 $\pm$ 0.42 (1.55 $\pm$ 0.45) &
  1.47 $\pm$ 0.29 (1.43 $\pm$ 0.38) &
  1.65 $\pm$ 0.47 (1.59 $\pm$ 0.46) \\
 &
  AR (SW) &
  2.30 $\pm$ 0.67 (2.24 $\pm$   0.72) &
  2.17 $\pm$ 0.84 (2.03 $\pm$ 0.81) &
  1.69 $\pm$ 0.48 (1.69 $\pm$ 0.50) &
  2.25 $\pm$ 0.72 (2.15 $\pm$ 0.75) \\
 &
  QS (SW) &
  2.02 $\pm$ 0.69 (1.87 $\pm$   0.61) &
  1.82 $\pm$ 0.55 (1.72 $\pm$ 0.47) &
  1.43 $\pm$ 0.33 (1.49 $\pm$ 0.51) &
  1.84 $\pm$ 0.62 (1.73 $\pm$ 0.56) \\
 &
  ALL (SW) &
  2.18 $\pm$ 0.68 (2.08 $\pm$   0.69) &
  1.85 $\pm$ 0.68 (1.74 $\pm$ 0.64) &
  1.50 $\pm$ 0.36 (1.50 $\pm$ 0.46) &
  2.00 $\pm$ 0.69 (1.88 $\pm$ 0.68) \\ \hline
\multirow{4}{*}{\textbf{$A_{He}$ {[}$\%${]}}} &
  CH (SW) &
  3.57 $\pm$ 1.79 (3.36 $\pm$   1.50) &
  3.11 $\pm$ 1.62 (3.10 $\pm$ 1.60) &
  2.48 $\pm$ 1.30 (2.55 $\pm$ 1.32) &
  3.19 $\pm$ 1.68 (3.11 $\pm$ 1.55) \\
 &
  AR (SW) &
  3.66 $\pm$ 2.52 (3.45 $\pm$   2.30) &
  3.01 $\pm$ 2.19 (3.25 $\pm$ 2.32) &
  2.24 $\pm$ 1.22 (2.66 $\pm$ 1.41) &
  3.45 $\pm$ 2.44 (3.35 $\pm$ 2.28) \\
 &
  QS (SW) &
  3.74 $\pm$ 2.24 (3.44 $\pm$   1.84) &
  3.48 $\pm$ 2.06 (3.37 $\pm$ 1.80) &
  2.31 $\pm$ 1.43 (2.57 $\pm$ 1.40) &
  3.43 $\pm$ 2.12 (3.24 $\pm$ 1.78) \\
 &
  ALL (SW) &
  3.65 $\pm$ 2.36 (3.42 $\pm$   2.07) &
  3.14 $\pm$ 1.92 (3.20 $\pm$ 1.90) &
  2.37 $\pm$ 1.33 (2.60 $\pm$ 1.37) &
  3.36 $\pm$ 2.18 (3.24 $\pm$ 1.95) \\ \hline
\multirow{4}{*}{\textbf{\begin{tabular}[c]{@{}c@{}}Magnetic field strength\\ (nT)\end{tabular}}} &
  CH (SW) &
  8.32 $\pm$ 2.95 (7.38 $\pm$   3.18) &
  8.32 $\pm$ 3.08 (7.18 $\pm$ 3.00) &
  7.98 $\pm$ 3.12 (5.94 $\pm$ 2.60) &
  8.27 $\pm$ 3.04 (7.05 $\pm$ 3.04) \\
 &
  AR (SW) &
  8.44 $\pm$ 3.39 (7.37 $\pm$   3.32) &
  8.18 $\pm$ 3.42 (7.00 $\pm$ 3.52) &
  7.27 $\pm$ 1.96 (5.73 $\pm$ 2.16) &
  8.34 $\pm$ 3.36 (7.19 $\pm$ 3.35) \\
 &
  QS (SW) &
  8.31 $\pm$ 3.26 (7.08 $\pm$   2.86) &
  7.69 $\pm$ 3.02 (6.17 $\pm$ 2.49) &
  6.95 $\pm$ 1.93 (5.07 $\pm$ 2.11) &
  7.84 $\pm$ 3.02 (6.30 $\pm$ 2.68) \\
 &
  ALL (SW) &
  8.40 $\pm$ 3.30 (7.29 $\pm$   3.22) &
  8.15 $\pm$ 3.21 (6.91 $\pm$ 3.13) &
  7.48 $\pm$ 2.57 (5.58 $\pm$ 2.38) &
  8.23 $\pm$ 3.22 (6.94 $\pm$ 3.14) \\ \hline
\multirow{4}{*}{\textbf{\begin{tabular}[c]{@{}c@{}}Proton Temperature\\ (10$^{5}$K)\end{tabular}}} &
  CH (SW) &
  1.05 $\pm$ 0.69 (1.24 $\pm$   0.81) &
  1.15 $\pm$ 0.70 (1.51 $\pm$ 0.87) &
  1.28 $\pm$ 0.70 (1.55 $\pm$ 0.78) &
  1.14 $\pm$ 0.70 (1.44 $\pm$ 0.85) \\
 &
  AR (SW) &
  0.85 $\pm$ 0.62 (1.09 $\pm$   0.82) &
  1.01 $\pm$ 0.97 (1.27 $\pm$ 0.97) &
  1.14 $\pm$ 0.73 (1.35 $\pm$ 0.85) &
  0.90 $\pm$ 0.73 (1.15 $\pm$ 0.87) \\
 &
  QS (SW) &
  0.89 $\pm$ 0.59 (1.01 $\pm$   0.61) &
  0.82 $\pm$ 0.48 (1.10 $\pm$ 0.65) &
  1.17 $\pm$ 0.68 (1.09 $\pm$ 0.61) &
  0.92 $\pm$ 0.58 (1.06 $\pm$ 0.63) \\
 &
  ALL (SW) &
  0.89 $\pm$ 0.63 (1.11 $\pm$   0.79) &
  1.04 $\pm$ 0.79 (1.36 $\pm$ 0.88) &
  1.21 $\pm$ 0.70 (1.34 $\pm$ 0.77) &
  0.97 $\pm$ 0.70 (1.24 $\pm$ 0.84) \\ \hline
\multirow{4}{*}{\textbf{\begin{tabular}[c]{@{}c@{}}Proton Number Density\\ (n/cc)\end{tabular}}} &
  CH (SW) &
  9.66 $\pm$ 6.91 (8.13 $\pm$   6.62) &
  9.09 $\pm$ 7.89 (6.76 $\pm$ 5.26) &
  9.62 $\pm$ 7.45 (7.61 $\pm$ 5.95) &
  9.33 $\pm$ 7.53 (7.30 $\pm$ 5.85) \\
 &
  AR (SW) &
  8.44 $\pm$ 7.68 (7.31 $\pm$   6.66) &
  7.81 $\pm$ 7.06 (6.06 $\pm$ 4.91) &
  6.97 $\pm$ 3.77 (7.05 $\pm$ 7.56) &
  8.25 $\pm$ 7.48 (6.96 $\pm$ 6.27) \\
 &
  QS (SW) &
  8.65 $\pm$ 7.43 (7.71 $\pm$   6.61) &
  7.90 $\pm$ 6.12 (6.64 $\pm$ 4.44) &
  11.78 $\pm$ 10.93 (6.81 $\pm$ 5.01) &
  8.52 $\pm$ 7.23 (7.21 $\pm$ 5.74) \\
 &
  ALL (SW) &
  8.69 $\pm$ 7.53 (7.39 $\pm$   6.47) &
  8.41 $\pm$ 7.32 (6.46 $\pm$ 5.00) &
  9.37 $\pm$ 7.78 (7.24 $\pm$ 6.23) &
  8.61 $\pm$ 7.47 (7.01 $\pm$ 5.92) \\ \hline
\multirow{4}{*}{\textbf{\begin{tabular}[c]{@{}c@{}}Speed\\ (km s$^{-1}$)\end{tabular}}} &
  CH (SW) &
  437 $\pm$ 80 (465 $\pm$ 101) &
  475 $\pm$ 93 (523 $\pm$ 110) &
  467 $\pm$ 88 (500 $\pm$ 115) &
  461 $\pm$ 90 (502 $\pm$ 111) \\
 &
  AR (SW) &
  410 $\pm$ 60 (430 $\pm$ 79) &
  436 $\pm$ 85 (459 $\pm$ 98) &
  446 $\pm$ 78 (480 $\pm$ 103) &
  418 $\pm$ 68 (441 $\pm$ 88) \\
 &
  QS (SW) &
  418 $\pm$ 64 (439 $\pm$ 75) &
  429 $\pm$ 60 (455 $\pm$ 82) &
  449 $\pm$ 69 (458 $\pm$ 84) &
  427 $\pm$ 65 (449 $\pm$ 80) \\
 &
  ALL (SW) &
  416 $\pm$ 65 (440 $\pm$ 86) &
  452 $\pm$ 87 (490 $\pm$ 108) &
  456 $\pm$ 81 (482 $\pm$ 105) &
  432 $\pm$ 77 (464 $\pm$ 100) \\ \hline
\multirow{4}{*}{\textbf{\begin{tabular}[c]{@{}c@{}}Expansion Velocity \tnote{f}\\ (km s$^{-1}$)\end{tabular}}} &
  CH &
  -0.06 $\pm$ 9.47 &
  -0.88 $\pm$ 9.73 &
  -0.77 $\pm$ 9.07 &
  -0.61 $\pm$ 9.55 \\
 &
  AR &
  0.14 $\pm$ 9.67 &
  -0.53 $\pm$ 9.18 &
  -0.60 $\pm$ 8.99 &
  -0.06 $\pm$ 9.52 \\
 &
  QS &
  0.08 $\pm$ 9.59 &
  -0.91 $\pm$ 8.71 &
  -3.03 $\pm$ 10.20 &
  -0.87 $\pm$ 9.42 \\
 &
  ALL &
  0.09 $\pm$ 9.61 &
  -0.77 $\pm$ 9.37 &
  -1.47 $\pm$ 9.43 &
  -0.38 $\pm$ 9.51 \\ \hline 
\multirow{4}{*}{\textbf{\begin{tabular}[c]{@{}c@{}}Correlation coefficient \\ between speed and O$^{7+}$/O$^{6+}$\end{tabular}}} &
  CH (SW) &
  -0.49 (-0.61) &
  -0.54 (-0.59) &
  -0.55 (-0.70) &
  -0.51 (-0.60) \\
 &
  AR (SW) &
  -0.37 (-0.37) &
  -0.29 (-0.40) &
  -0.63 (-0.73) &
  -0.36 (-0.40) \\
 &
  QS (SW) &
  -0.14 (-0.24) &
  -0.20 (-0.47) &
  -0.60 (-0.61) &
  -0.20 (-0.34) \\
 &
  ALL (SW) &
  -0.36 (-0.41) &
  -0.43 (-0.53) &
  -0.53 (-0.67) &
  -0.40 (-0.47) \\ \hline
 &
   &
   &
   &
   &
  
\end{tabular}%
  \begin{tablenotes}
%\item[ ] \textbf{The values presented in the table are the average and standard derivations of the parameters inside SFRs and background solar wind.}
\item[a] Average duration of SFRs.
\item[b] Background solar wind of SFRs.
\item[c] The O$^{7+}$/O$^{6+}$ (C$^{6+}$/C$^{5+}$, C$^{6+}$/C$^{4+}$) represents the density ratios between O$^{7+}$ and O$^{6+}$ (C$^{6+}$ and C$^{5+}$, C$^{6+}$ and C$^{4+}$), and the values presented in the table are the average and standard derivations of the parameters inside SFRs and background solar wind.
\item[d] The FIP bias is derived from the ratios between Fe/O detected in-situ and their corresponding photospheric values of 0.065 \citep{2009ARA&A..47..481A}.
\item[e] The FIP bias is derived from the ratios between Mg/O detected in-situ and their corresponding photospheric values of 0.081 
\citep {2009ARA&A..47..481A}.
\item[f] The expansion velocity of SFRs is derived from the half of speed difference between the leading and trailing edges of SFRs \citep{ 2016JGRA..121.5005Y,2018ApJS..239...12H}.
   \end{tablenotes}      
    \end{threeparttable} 
}
\end{table}

\begin{figure}
\centering
\includegraphics[width=1.0\textwidth]{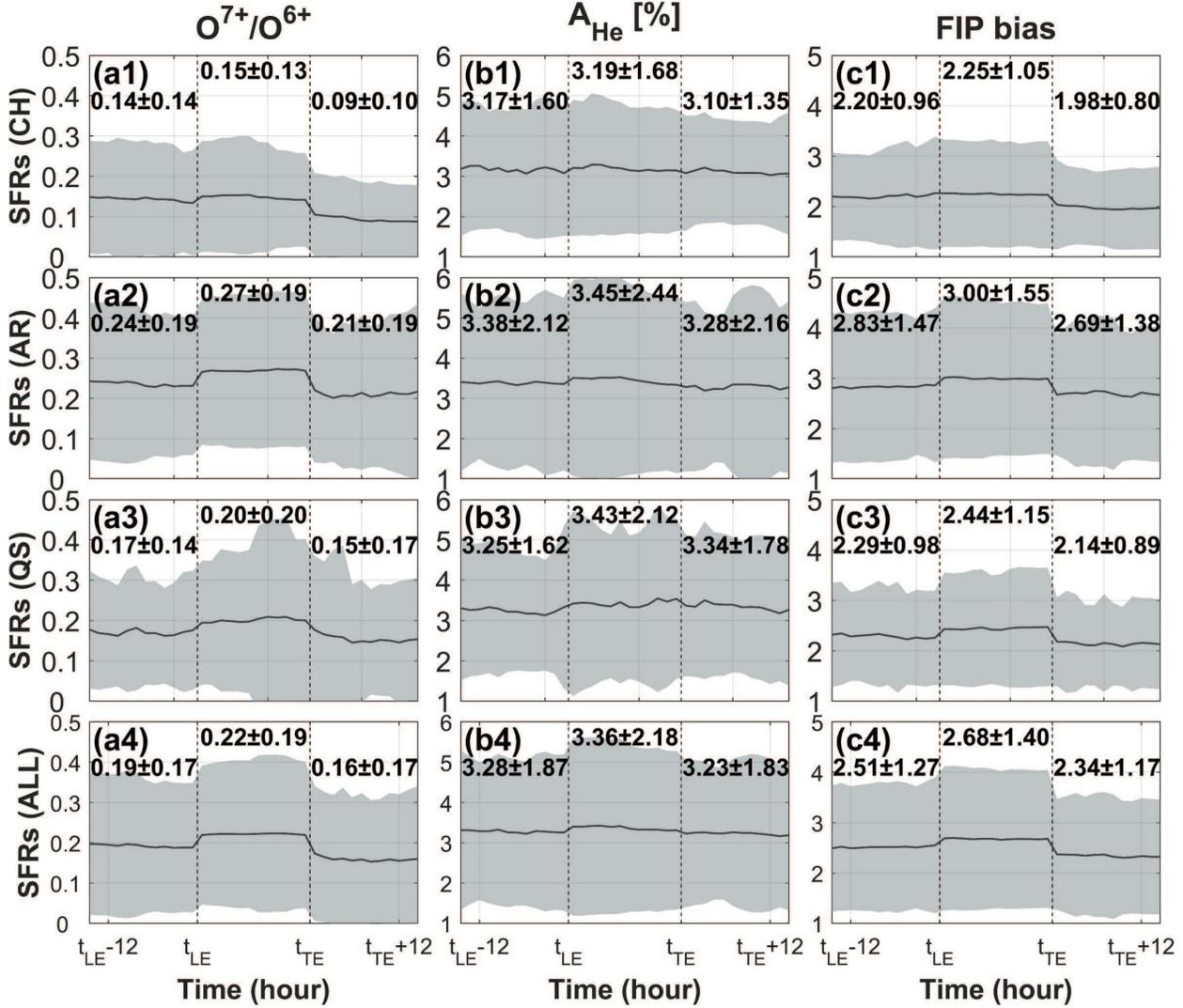}
\caption{The compositional signatures of SFRs and their background in three types of solar wind. The O$^{7+}$/O$^{6+}$ (left column), helium abundance (middle column), and FIP bias (derived from Fe/O, right column) inside CH, AR, QS solar wind, and SFRs as a whole are shown in the first to fourth rows. The duration of each SFR is normalized to 12 hours, which is equal to the duration of background solar wind before and after SFRs. The leading edges (LE) and trailing edges (TE) of SFRs are denoted by vertical dashed lines. The black lines show the average values and the 1-sigma standard deviations {as} demonstrated by shaded areas. The average values inside SFRs and associated background solar wind are also given in the panels.}
\label{fig:1}
\end{figure}

The total number of SFRs (duration ranging from 1 to 12 hours) that are analyzed in the present study is 1993. The numbers of SFRs that are immersed in the identified CH, AR, and QS solar wind are 652, 988, and 353, respectively. The properties of SFRs immersed in CH, AR, and QS solar wind and their associated background are summarized and listed in Table~\ref{tab:1}. The parameters are given according to solar activity phases, considering that the properties of SFRs \citep{2018ApJS..239...12H} and solar wind \citep{2009GeoRL..3614104Z,2015SoPh..290.1399F} both change with solar activities.
Generally, the properties of SFRs immersed in CH, AR, and QS solar wind are significantly different from each other. There are only small differences between the parameters inside SFRs and their associated background in all three types of solar wind (see Table~\ref{tab:1}).

The compositional signatures of SFRs and their background immersed in CH, AR, and QS solar wind are presented in Figure~\ref{fig:1}. 
%It should be noted that the values presented in \cref{fig:1,fig:3,fig:4} have small discrepancies from those shown in the associated tables. The values presented in the tables represent the average and standard derivations of the parameters inside SFRs and background solar wind. However, the durations of SFRs shown in the figures are normalized to 12 hours, which is equal to the duration of background solar wind before and after SFRs. The above display method makes the comparison between SFRs and background solar wind more clear in the figures. Qualitatively, the relative relationships of the parameters between the SFRs and background solar are the same in the figures and associated tables. Although the absolute values of the parameters are not exactly equal. The parameters of background solar wind in the tables and figures are also not the same. As the values in the tables represent the average and standard derivations of background solar wind before and after SFRs together. The values of background solar wind before and after SFRs are presented separately in the figures.}
The O$^{7+}$/O$^{6+}$ is the lowest (highest) in CH (AR) SFRs, with QS SFRs lying in between. The O$^{7+}$/O$^{6+}$ of AR SFRs is about twice as that of CH SFRs. The O$^{7+}$/O$^{6+}$ inside all three types of SFRs is slightly higher than that in the background solar wind. Generally, the O$^{7+}$/O$^{6+}$ inside SFRs is 10$\%$-20$\%$ higher than that in the background solar wind. The O$^{7+}$/O$^{6+}$ inside CH, AR, and QS SFRs (background) is 0.15 $\pm$ 0.13 (0.12 $\pm$ 0.12), 0.27 $\pm$ 0.19 (0.22 $\pm$ 0.19), and 0.20 $\pm$ 0.20 (0.16 $\pm$ 0.16), respectively. 
The charge states of C$^{6+}$/C$^{5+}$ and C$^{6+}$/C$^{4+}$ are also presented in Table~\ref{tab:1}. Qualitatively, the characteristics of C$^{6+}$/C$^{5+}$ and C$^{6+}$/C$^{4+}$ inside SFRs and background solar wind are similar to those of O$^{7+}$/O$^{6+}$.
The characteristics of FIP bias are almost the same as those of O$^{7+}$/O$^{6+}$. The $A_{He}$ inside SFRs is also slightly higher than that in the associated background solar wind. It is worth noting that the $A_{He}$ in three types of solar wind is only slightly different from each other. The clear difference is that the $A_{He}$ range is wider inside AR and QS SFRs than that in CH SFRs.
Generally, the $A_{He}$ is higher (4$\%$-5$\%$) and uniform in the fast solar wind \citep{2008GeoRL..3518103M}. The distribution range is wider and the average value is lower for the $A_{He}$ inside the slow solar wind \citep{2001GeoRL..28.2767A,2007ApJ...660..901K,2012ApJ...745..162K,2018MNRAS.478.1884F}. 
The typical fast solar wind originating from large polar coronal holes is generally associated with high $A_{He}$. The speed of typical fast solar wind is generally higher than 600 km s$^{-1}$, and the average speed is about 700 km s$^{-1}$ (Figure~2 in \citet{2008GeoRL..3518103M}). However, the above typical fast solar wind rarely appears near the Earth, considering that the CH solar wind in the ecliptic plane generally originates from small, mid-scale CHs, and/or boundaries of CHs. The average speed of AR and QS solar wind (about 450 km s$^{-1}$) is only slightly lower than that of CH solar wind (about 500 km s$^{-1}$, see top panel of Figure~5 in \citet{2015SoPh..290.1399F}). Therefore, the $A_{He}$ inside CH wind in the ecliptic plane is lower than that of typical fast solar wind originating from large polar coronal holes. The $A_{He}$ inside CH wind is not higher than that in AR and QS solar wind in the ecliptic plane.

We suggest that the classification procedure for three types of solar wind in \citet{2015SoPh..290.1399F} is reasonable. The statistical results demonstrated that the O$^{7+}$/O$^{6+}$ in AR solar wind is about twice as the O$^{7+}$/O$^{6+}$ in CH wind (top panel of Figure 5 in \citet{2015SoPh..290.1399F}), although there is no significant difference between the speed of CH and AR solar wind. This is consistent with the fact that the temperatures of source regions for AR solar wind are significantly higher than those for CH solar wind. It is worth noting that the compositional signatures of SFRs are higher than those of the background in all three types of solar wind. The average values of O$^{7+}$/O$^{6+}$ inside SFRs are all ~10$\%$-20$\%$ higher than those in the associated background in all three types of solar wind. Is the slightly higher O$^{7+}$/O$^{6+}$ inside SFRs a common feature in all SFRs? Or is it caused by significantly higher O$^{7+}$/O$^{6+}$ only in part of SFRs? Are there also part of SFRs with significantly lower O$^{7+}$/O$^{6+}$? In the next section, we will investigate and analyze the compositional signatures of SFRs in more details.

\subsection{The comparisons between the hot-SFRs, cold-SFRs, normal-SFRs, and their associated background immersed in CH, AR, and QS solar wind}

\begin{table}[]
\centering
\caption{The paramters of hot-SFRs and background in CH, AR, and QS solar wind}
\label{tab:2}
\resizebox{\columnwidth}{!}{%
\begin{threeparttable} 
\begin{tabular}{cccccc}
\hline
 &
   &
  \textbf{MAX} &
  \textbf{DEC} &
  \textbf{MIN} &
  \textbf{TOTAL} \\ \hline
\multirow{4}{*}{\textbf{Number of SFRs}} &
  CH &
  32 &
  65 &
  26 &
  123 \\
 &
  AR &
  106 &
  34 &
  6 &
  146 \\
 &
  QS &
  22 &
  26 &
  4 &
  52 \\
 &
  ALL &
  160 &
  125 &
  36 &
  321 \\ \hline
\multirow{4}{*}{\textbf{Duration (hour) \tnote{a}}} &
  CH &
  2.09 $\pm$ 1.44 &
  2.08 $\pm$ 1.52 &
  1.96 $\pm$ 0.96 &
  2.06 $\pm$ 1.39 \\
 &
  AR &
  2.51 $\pm$ 2.09 &
  2.32 $\pm$ 1.43 &
  2.50 $\pm$ 1.52 &
  2.47 $\pm$ 1.93 \\
 &
  QS &
  2.41 $\pm$ 2.11 &
  2.50 $\pm$ 1.70 &
  2.75 $\pm$ 1.50 &
  2.48 $\pm$ 1.84 \\
 &
  ALL &
  2.41 $\pm$ 1.98 &
  2.23 $\pm$ 1.54 &
  2.14 $\pm$ 1.13 &
  2.31 $\pm$ 1.73 \\ \hline
\multirow{4}{*}{\textbf{O$^{7+}$/O$^{6+}$ \tnote{c}}} &
  CH (SW)\tnote{b} &
  0.37 $\pm$   0.23 (0.21 $\pm$ 0.17) &
  0.23 $\pm$ 0.17 (0.11 $\pm$ 0.12) &
  0.13 $\pm$ 0.08 (0.05 $\pm$ 0.06) &
  0.24 $\pm$ 0.20 (0.13 $\pm$ 0.14) \\
 &
  AR (SW) &
  0.47 $\pm$ 0.26 (0.26 $\pm$   0.21) &
  0.31 $\pm$ 0.24 (0.17 $\pm$ 0.15) &
  0.17 $\pm$ 0.10 (0.08 $\pm$ 0.08) &
  0.43 $\pm$ 0.27 (0.23 $\pm$ 0.20) \\
 &
  QS (SW) &
  0.48 $\pm$ 0.42 (0.25 $\pm$   0.18) &
  0.24 $\pm$ 0.10 (0.13 $\pm$ 0.08) &
  0.06 $\pm$ 0.05 (0.05 $\pm$ 0.05) &
  0.32 $\pm$ 0.31 (0.17 $\pm$ 0.15) \\
 &
  ALL (SW) &
  0.46 $\pm$ 0.29 (0.25 $\pm$   0.20) &
  0.25 $\pm$ 0.18 (0.13 $\pm$ 0.12) &
  0.13 $\pm$ 0.08 (0.06 $\pm$ 0.06) &
  0.35 $\pm$ 0.27 (0.18 $\pm$ 0.18) \\ \hline
\multirow{4}{*}{\textbf{C$^{6+}$/C$^{5+}$ \tnote{c}}} &
  CH (SW) &
  1.17 $\pm$ 0.80 (0.97 $\pm$   0.64) &
  1.03 $\pm$ 0.67 (0.65 $\pm$ 0.56) &
  0.81 $\pm$ 0.43 (0.48 $\pm$ 0.40) &
  1.02 $\pm$ 0.68 (0.71 $\pm$ 0.58) \\
 &
  AR (SW) &
  1.38 $\pm$ 0.86 (1.12 $\pm$   0.78) &
  1.44 $\pm$ 1.03 (0.97 $\pm$ 0.59) &
  1.08 $\pm$ 0.57 (0.61 $\pm$ 0.49) &
  1.38 $\pm$ 0.89 (1.06 $\pm$ 0.73) \\
 &
  QS (SW) &
  1.35 $\pm$ 0.89 (1.21 $\pm$   0.71) &
  1.23 $\pm$ 0.62 (0.81 $\pm$ 0.46) &
  0.61 $\pm$ 0.51 (0.51 $\pm$ 0.38) &
  1.22 $\pm$ 0.76 (0.94 $\pm$ 0.62) \\
 &
  ALL (SW) &
  1.34 $\pm$ 0.86 (1.10 $\pm$   0.74) &
  1.19 $\pm$ 0.80 (0.77 $\pm$ 0.56) &
  0.83 $\pm$ 0.48 (0.51 $\pm$ 0.41) &
  1.23 $\pm$ 0.82 (0.90 $\pm$ 0.68) \\ \hline
\multirow{4}{*}{\textbf{C$^{6+}$/C$^{4+}$ \tnote{c}}} &
  CH (SW) &
  4.67 $\pm$ 4.03 (3.52 $\pm$   3.08) &
  3.78 $\pm$ 2.99 (2.12 $\pm$ 2.27) &
  2.34 $\pm$ 2.12 (1.07 $\pm$ 1.23) &
  3.73 $\pm$ 3.24 (2.30 $\pm$ 2.51) \\
 &
  AR (SW) &
  5.85 $\pm$ 5.13 (3.94 $\pm$   3.50) &
  5.42 $\pm$ 4.26 (3.19 $\pm$ 2.56) &
  3.44 $\pm$ 2.89 (1.55 $\pm$ 1.66) &
  5.66 $\pm$ 4.89 (3.65 $\pm$ 3.28) \\
 &
  QS (SW) &
  5.23 $\pm$ 3.46 (4.46 $\pm$   3.55) &
  4.35 $\pm$ 3.41 (2.57 $\pm$ 1.87) &
  1.34 $\pm$ 1.10 (1.11 $\pm$ 1.13) &
  4.46 $\pm$ 3.44 (3.19 $\pm$ 2.89) \\
 &
  ALL (SW) &
  5.56 $\pm$ 4.76 (3.95 $\pm$   3.43) &
  4.38 $\pm$ 3.55 (2.49 $\pm$ 2.32) &
  2.41 $\pm$ 2.24 (1.16 $\pm$ 1.31) &
  4.79 $\pm$ 4.24 (3.05 $\pm$ 3.00) \\ \hline
\multirow{4}{*}{\textbf{FIP bias (Fe/O) \tnote{d}}} &
  CH (SW) &
  2.85 $\pm$ 1.81 (2.34 $\pm$   1.01) &
  2.04 $\pm$ 0.84 (1.92 $\pm$ 0.76) &
  2.60 $\pm$ 1.60 (1.96 $\pm$ 0.97) &
  2.37 $\pm$ 1.37 (2.04 $\pm$ 0.89) \\
 &
  AR (SW) &
  3.42 $\pm$ 1.95 (2.94 $\pm$   1.69) &
  2.30 $\pm$ 1.09 (2.30 $\pm$ 1.03) &
  2.17 $\pm$ 0.83 (2.13 $\pm$ 0.70) &
  3.13 $\pm$ 1.83 (2.75 $\pm$ 1.56) \\
 &
  QS (SW) &
  2.67 $\pm$ 1.15 (2.37 $\pm$   1.06) &
  2.26 $\pm$ 0.67 (2.07 $\pm$ 0.70) &
  1.34 $\pm$ 0.62 (1.65 $\pm$ 0.60) &
  2.35 $\pm$ 0.96 (2.15 $\pm$ 0.89) \\
 &
  ALL (SW) &
  3.22 $\pm$ 1.86 (2.74 $\pm$   1.51) &
  2.17 $\pm$ 0.89 (2.05 $\pm$ 0.84) &
  2.34 $\pm$ 1.44 (1.93 $\pm$ 0.88) &
  2.73 $\pm$ 1.60 (2.37 $\pm$ 1.27) \\ \hline
\multirow{4}{*}{\textbf{FIP bias (Mg/O) \tnote{e}}} &
  CH (SW) &
  2.12 $\pm$ 0.90 (1.79 $\pm$   0.51) &
  1.70 $\pm$ 0.47 (1.50 $\pm$ 0.38) &
  1.55 $\pm$ 0.34 (1.45 $\pm$ 0.37) &
  1.78 $\pm$ 0.63 (1.57 $\pm$ 0.44) \\
 &
  AR (SW) &
  2.57 $\pm$ 0.84 (2.30 $\pm$   0.78) &
  2.06 $\pm$ 0.63 (1.94 $\pm$ 0.63) &
  1.78 $\pm$ 0.42 (1.69 $\pm$ 0.43) &
  2.43 $\pm$ 0.82 (2.19 $\pm$ 0.76) \\
 &
  QS (SW) &
  2.30 $\pm$ 0.85 (2.02 $\pm$   0.76) &
  1.68 $\pm$ 0.39 (1.65 $\pm$ 0.45) &
  1.48 $\pm$ 0.37 (1.42 $\pm$ 0.29) &
  1.92 $\pm$ 0.70 (1.78 $\pm$ 0.63) \\
 &
  ALL (SW) &
  2.46 $\pm$ 0.87 (2.16 $\pm$   0.76) &
  1.80 $\pm$ 0.53 (1.64 $\pm$ 0.51) &
  1.58 $\pm$ 0.37 (1.49 $\pm$ 0.38) &
  2.12 $\pm$ 0.80 (1.88 $\pm$ 0.69) \\ \hline
\multirow{4}{*}{\textbf{$A_{He}$ {[}$\%${]}}} &
  CH (SW) &
  4.03 $\pm$ 2.57 (3.39 $\pm$   1.71) &
  2.94 $\pm$ 1.72 (3.01 $\pm$ 1.53) &
  2.09 $\pm$ 1.24 (2.72 $\pm$ 1.46) &
  3.13 $\pm$ 2.06 (3.07 $\pm$ 1.59) \\
 &
  AR (SW) &
  3.91 $\pm$ 2.48 (3.46 $\pm$   2.53) &
  3.65 $\pm$ 3.72 (2.71 $\pm$ 2.17) &
  1.78 $\pm$ 0.67 (2.54 $\pm$ 1.12) &
  3.80 $\pm$ 2.79 (3.23 $\pm$ 2.43) \\
 &
  QS (SW) &
  4.52 $\pm$ 3.60 (3.25 $\pm$   1.93) &
  4.01 $\pm$ 2.05 (3.49 $\pm$ 1.61) &
  2.43 $\pm$ 0.66 (3.06 $\pm$ 1.52) &
  4.16 $\pm$ 2.84 (3.35 $\pm$ 1.75) \\
 &
  ALL (SW) &
  4.01 $\pm$ 2.67 (3.41 $\pm$   2.28) &
  3.36 $\pm$ 2.52 (3.03 $\pm$ 1.74) &
  2.07 $\pm$ 1.14 (2.77 $\pm$ 1.44) &
  3.62 $\pm$ 2.59 (3.19 $\pm$ 2.00) \\ \hline
\multirow{4}{*}{\textbf{\begin{tabular}[c]{@{}c@{}}Magnetic field strength\\ (nT)\end{tabular}}} &
  CH (SW) &
  9.53 $\pm$ 3.27 (7.76 $\pm$   3.61) &
  8.56 $\pm$ 3.02 (7.61 $\pm$ 3.33) &
  8.04 $\pm$ 2.70 (6.31 $\pm$ 2.60) &
  8.74 $\pm$ 3.09 (7.42 $\pm$ 3.34) \\
 &
  AR (SW) &
  9.03 $\pm$ 3.55 (8.10 $\pm$   3.65) &
  7.82 $\pm$ 3.20 (6.47 $\pm$ 2.63) &
  7.93 $\pm$ 1.66 (6.18 $\pm$ 1.91) &
  8.73 $\pm$ 3.46 (7.61 $\pm$ 3.44) \\
 &
  QS (SW) &
  9.77 $\pm$ 3.64 (7.50 $\pm$   3.64) &
  9.84 $\pm$ 3.85 (7.03 $\pm$ 3.05) &
  7.52 $\pm$ 2.07 (4.48 $\pm$ 1.42) &
  9.59 $\pm$ 3.68 (6.91 $\pm$ 3.32) \\
 &
  ALL (SW) &
  9.22 $\pm$ 3.52 (7.93 $\pm$   3.66) &
  8.60 $\pm$ 3.33 (7.17 $\pm$ 3.12) &
  7.94 $\pm$ 2.43 (5.91 $\pm$ 2.37) &
  8.87 $\pm$ 3.39 (7.40 $\pm$ 3.39) \\ \hline
\multirow{4}{*}{\textbf{\begin{tabular}[c]{@{}c@{}}Proton Temperature\\ (10$^{5}$K)\end{tabular}}} &
  CH (SW) &
  0.79 $\pm$ 0.48 (1.22 $\pm$   0.94) &
  0.99 $\pm$ 0.62 (1.59 $\pm$ 0.99) &
  1.06 $\pm$ 0.65 (1.50 $\pm$ 0.73) &
  0.94 $\pm$ 0.60 (1.48 $\pm$ 0.95) \\
 &
  AR (SW) &
  0.71 $\pm$ 0.65 (1.21 $\pm$   0.95) &
  0.86 $\pm$ 0.48 (1.22 $\pm$ 0.80) &
  1.30 $\pm$ 0.63 (1.58 $\pm$ 0.93) &
  0.76 $\pm$ 0.63 (1.23 $\pm$ 0.92) \\
 &
  QS (SW) &
  0.70 $\pm$ 0.44 (0.93 $\pm$   0.53) &
  0.83 $\pm$ 0.50 (1.14 $\pm$ 0.70) &
  0.89 $\pm$ 0.19 (0.99 $\pm$ 0.49) &
  0.78 $\pm$ 0.46 (1.04 $\pm$ 0.62) \\
 &
  ALL (SW) &
  0.73 $\pm$ 0.60 (1.16 $\pm$   0.90) &
  0.92 $\pm$ 0.57 (1.40 $\pm$ 0.90) &
  1.07 $\pm$ 0.60 (1.41 $\pm$ 0.76) &
  0.83 $\pm$ 0.60 (1.28 $\pm$ 0.90) \\ \hline
\multirow{4}{*}{\textbf{\begin{tabular}[c]{@{}c@{}}Proton Number Density\\ (n/cc)\end{tabular}}} &
  CH (SW) &
  13.91 $\pm$ 8.51 (10.13 $\pm$   9.06) &
  11.86 $\pm$ 9.34 (8.55 $\pm$ 7.05) &
  9.38 $\pm$ 6.09 (8.58 $\pm$ 7.38) &
  12.23 $\pm$ 8.88 (9.03 $\pm$ 7.77) \\
 &
  AR (SW) &
  9.84 $\pm$ 8.98 (7.97 $\pm$   6.74) &
  8.23 $\pm$ 6.03 (6.27 $\pm$ 4.56) &
  4.58 $\pm$ 2.48 (4.46 $\pm$ 2.96) &
  9.43 $\pm$ 8.43 (7.52 $\pm$ 6.31) \\
 &
  QS (SW) &
  11.43 $\pm$ 7.73 (9.45 $\pm$   8.60) &
  7.37 $\pm$ 6.76 (6.76 $\pm$ 4.44) &
  10.84 $\pm$ 1.90 (8.14 $\pm$ 2.67) &
  9.22 $\pm$ 7.50 (8.13 $\pm$ 6.97) \\
 &
  ALL (SW) &
  10.80 $\pm$ 8.89 (8.65 $\pm$   7.64) &
  9.88 $\pm$ 8.25 (7.56 $\pm$ 5.96) &
  8.66 $\pm$ 5.94 (7.68 $\pm$ 6.71) &
  10.35 $\pm$ 8.56 (8.16 $\pm$ 6.98) \\ \hline
\multirow{4}{*}{\textbf{\begin{tabular}[c]{@{}c@{}}Speed\\ (km s$^{-1}$)\end{tabular}}} &
  CH (SW) &
  414 $\pm$ 63 (440 $\pm$ 83) &
  455 $\pm$ 98 (519 $\pm$ 121) &
  457 $\pm$ 88 (493 $\pm$ 100) &
  444 $\pm$ 89 (493 $\pm$ 114) \\
 &
  AR (SW) &
  409 $\pm$ 60 (436 $\pm$ 77) &
  437 $\pm$ 82 (448 $\pm$ 82) &
  436 $\pm$ 61 (465 $\pm$ 109) &
  417 $\pm$ 66 (440 $\pm$ 81) \\
 &
  QS (SW) &
  406 $\pm$ 66 (428 $\pm$ 67) &
  452 $\pm$ 54 (474 $\pm$ 65) &
  425 $\pm$ 43 (448 $\pm$ 59) &
  429 $\pm$ 63 (451 $\pm$ 69) \\
 &
  ALL (SW) &
  410 $\pm$ 61 (435 $\pm$ 77) &
  449 $\pm$ 86 (491 $\pm$ 107) &
  447 $\pm$ 78 (478 $\pm$ 97) &
  428 $\pm$ 75 (462 $\pm$ 96) \\ \hline
\multirow{4}{*}{\textbf{\begin{tabular}[c]{@{}c@{}}Expansion Velocity \tnote{f}\\ (km s$^{-1}$)\end{tabular}}} &
  CH &
  0.63 $\pm$ 10.14 &
  -0.80 $\pm$ 10.06 &
  -0.99 $\pm$ 9.16 &
  -0.61 $\pm$ 9.90 \\
 &
  AR &
  0.48 $\pm$ 9.77 &
  -0.43 $\pm$ 8.78 &
  -0.58 $\pm$ 8.38 &
  -0.02 $\pm$ 9.23 \\
 &
  QS &
  -0.33 $\pm$ 9.56 &
  -1.16 $\pm$ 8.45 &
  -3.34 $\pm$ 9.90 &
  -1.48 $\pm$ 9.14 \\
 &
  ALL &
  0.39 $\pm$ 9.80 &
  -0.73 $\pm$ 9.36 &
  -1.61 $\pm$ 9.28 &
  -0.53 $\pm$ 9.50 \\ \hline
\multirow{4}{*}{\textbf{\begin{tabular}[c]{@{}c@{}}Correlation coefficient\\ between speed and O$^{7+}$/O$^{6+}$\end{tabular}}} &
  CH (SW) &
  -0.59 (-0.56) &
  -0.38 (-0.52) &
  -0.66 (-0.71) &
  -0.46 (-0.54) \\
 &
  AR (SW) &
  -0.31 (-0.43) &
  -0.08 (-0.33) &
  -0.56 (-0.71) &
  -0.29 (-0.42) \\
 &
  QS (SW) &
  0.14 (-0.19) &
  -0.29 (-0.51) &
  -0.73 (-0.79) &
  -0.12 (-0.35) \\
 &
  ALL (SW) &
  -0.26 (-0.42) &
  -0.27 (-0.48) &
  -0.56 (-0.69) &
  -0.34 (-0.47) \\ \hline
 &
   &
   &
   &
   &
  
\end{tabular}%
  \begin{tablenotes}
%\item[ ] \textbf{The values presented in the table are the average and standard derivations of the parameters inside SFRs and background solar wind.}
\item[a] Average duration of SFRs.
\item[b] Background solar wind of SFRs.
\item[c] The O$^{7+}$/O$^{6+}$ (C$^{6+}$/C$^{5+}$, C$^{6+}$/C$^{4+}$) represents the density ratios between O$^{7+}$ and O$^{6+}$ (C$^{6+}$ and C$^{5+}$, C$^{6+}$ and C$^{4+}$), and the values presented in the table are the average and standard derivations of the parameters inside SFRs and background solar wind.
\item[d] The FIP bias is derived from the ratios between Fe/O detected in-situ and their corresponding photospheric values of 0.065 \citep{2009ARA&A..47..481A}.
\item[e] The FIP bias is derived from the ratios between Mg/O detected in-situ and their corresponding photospheric values of 0.081 
\citep {2009ARA&A..47..481A}.
\item[f] The expansion velocity of SFRs is derived from the half of speed difference between the leading and trailing edges of SFRs \citep{ 2016JGRA..121.5005Y,2018ApJS..239...12H}.
   \end{tablenotes}      
    \end{threeparttable} 
}
\end{table}

\begin{figure}
\centering
\includegraphics[width=1.0\textwidth]{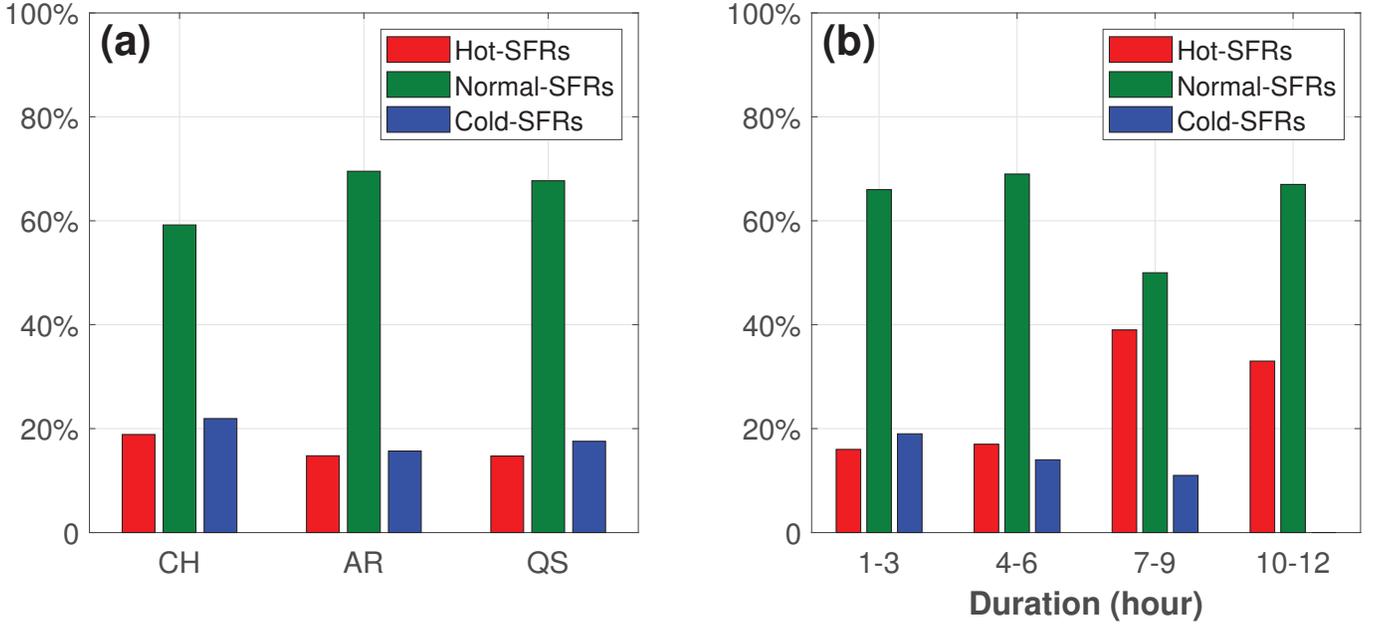}
\caption{The proportions of hot-SFRs, cold-SFRs, and normal-SFRs inside three types of solar wind (left panel) and proportion variations with duration of hot-SFRs, cold-SFRs, and normal-SFRs (right panel) are presented. The red, blue, and green represent hot-SFRs, cold-SFRs, and normal-SFRs, respectively. The significant characteristic  is that the proportions of hot-SFRs (cold-SFRs) increase (decrease) with increasing of duration.}
\label{fig:2}
\end{figure}

\begin{figure}
\centering
\includegraphics[width=1.0\textwidth]{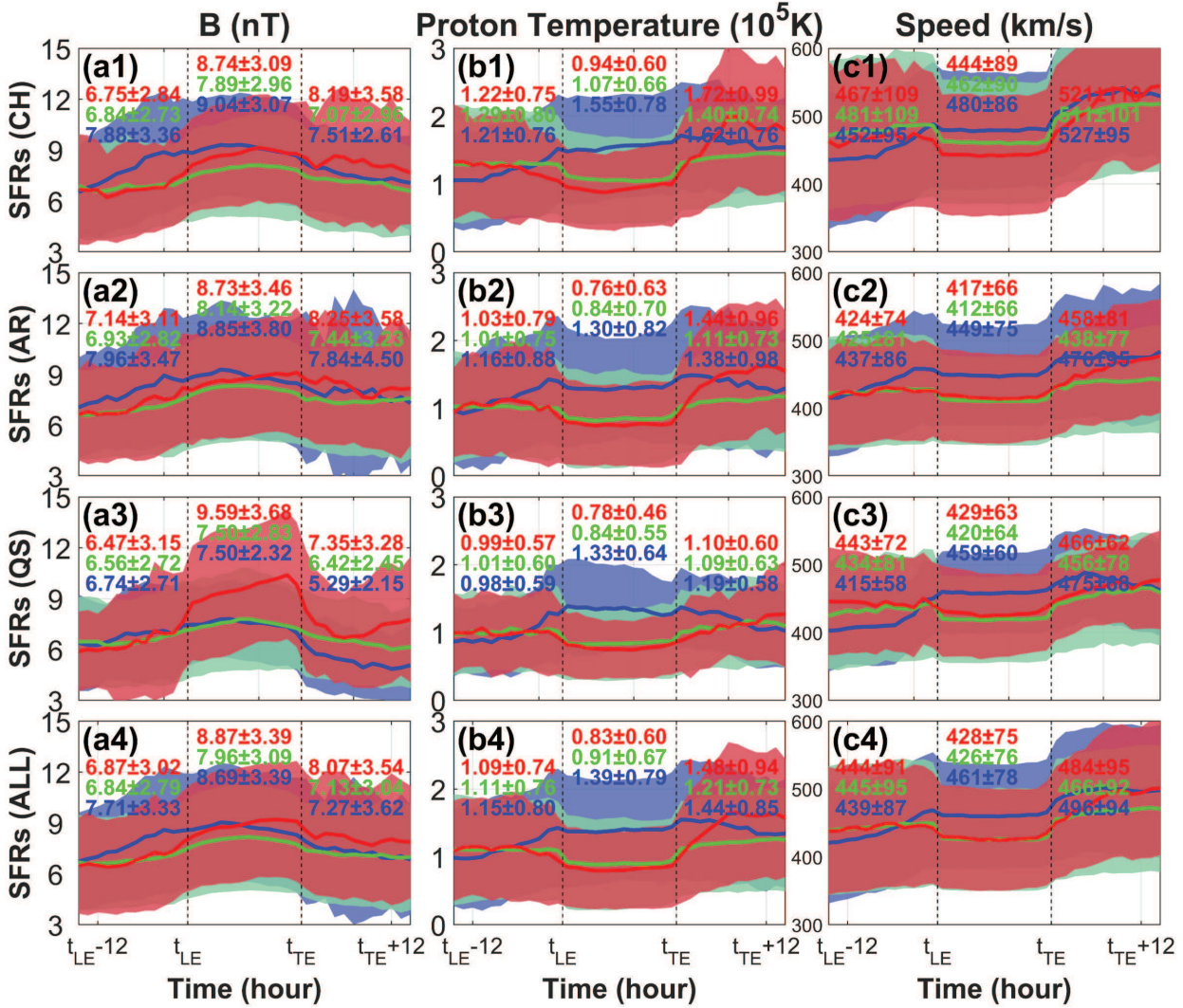}
\caption{The magnetic field strengths, proton temperatures, and speeds of hot-SFRs, cold-SFRs, normal-SFRs, and their associated background solar wind are shown from left to right. The SFRs inside CH, AR, and QS solar wind, and SFRs as a whole are presented from the top to the bottom. The duration of each SFR is normalized to 12 hours, which is equal to the duration of background solar wind before and after SFRs. The leading edges (LE) and trailing edges (TE) of SFRs are denoted by vertical dashed lines. The red, blue, and green represent hot-SFRs, cold-SFRs, and normal-SFRs. The solid lines denote the average values and the shaded areas represent the 1-sigma standard deviations. The average values inside hot-SFRs, cold-SFRs, normal-SFRs, and associated background solar wind are also given in each panel.}
\label{fig:3}
\end{figure}

\begin{figure}
\centering
\includegraphics[width=1.0\textwidth]{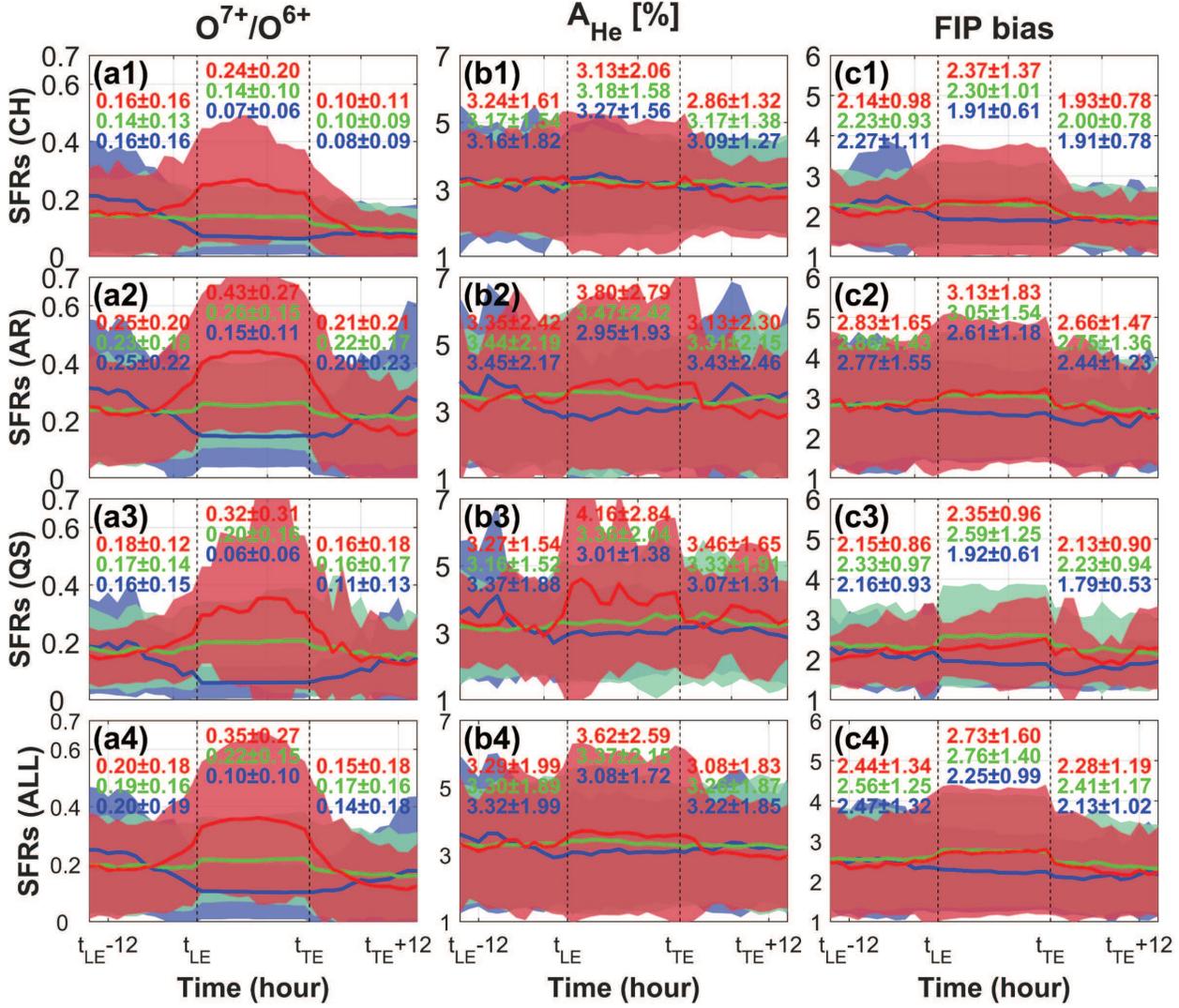}
\caption{The compositional signatures inside hot-SFRs, cold-SFRs, normal-SFRs, and associated background solar wind. The O$^{7+}$/O$^{6+}$, helium abundance, and FIP bias (derived from Fe/O) are presented from left to right. The SFRs inside CH, AR, and QS solar wind, and SFRs as a whole are presented from the top to the bottom. The duration of each SFR is normalized to 12 hours, which is equal to the duration of background solar wind before and after SFRs. The leading edges (LE) and trailing edges (TE) of SFRs are denoted by vertical dashed lines. The red, blue,and green represent hot-SFRs, cold-SFRs, and normal-SFRs. The solid lines denote the average values and the shaded areas represent the 1-sigma standard deviations. The average values inside hot-SFRs, cold-SFRs, normal-SFRs, and associated background solar wind are also given in each panel.}
\label{fig:4}
\end{figure}

\begin{figure}
\centering
\includegraphics[width=1.0\textwidth]{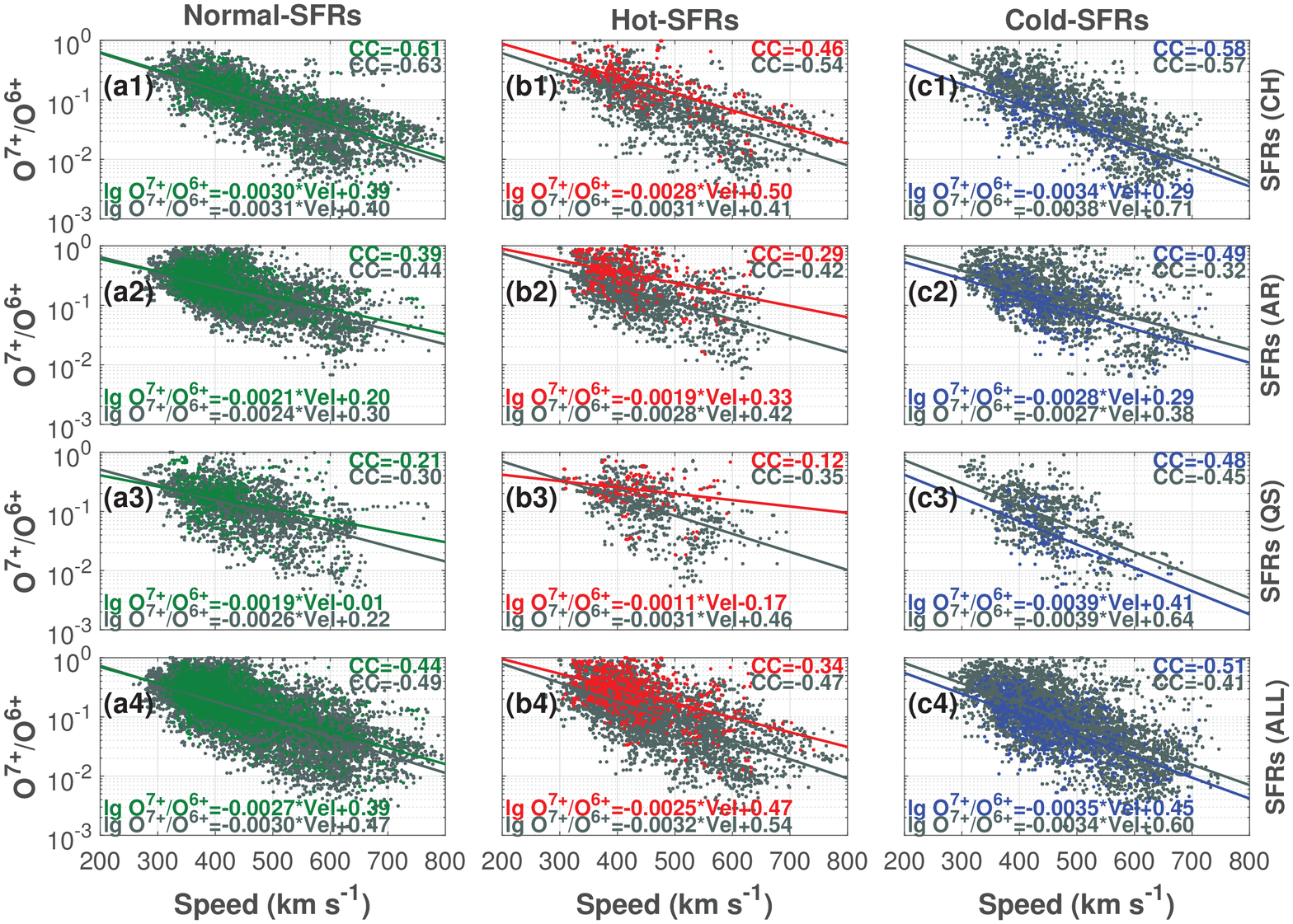}
\caption{The distributions in speed versus O$^{7+}$/O$^{6+}$ space for normal-SFRs (left panels), hot-SFRs (middle panels), and cold-SFRs (right panels) immersed in CH, AR, QS, and solar wind as a whole (from top to the bottom). The green, red, and blue represent normal-SFRs, hot-SFRs, and cold-SFRs, respectively. The distributions of background solar wind (grey) of hot-SFRs, cold-SFRS, and normal-SFRs are also presented for comparison. The correlation coefficients (CCs) and linear fitting results between the speed and  O$^{7+}$/O$^{6+}$ are also given in each panel.} 
\label{fig:5}
\end{figure}

To explore the characteristics of SFRs further, we classify the SFRs into hot-SFRs, cold-SFRs, and normal-SFRs according to whether or not the O$^{7+}$/O$^{6+}$ is clearly elevated or dropped inside SFRs. An SFR is defined as hot-SFR (cold-SFRs) if the O$^{7+}$/O$^{6+}$ inside the SFR is 30$\%$ higher (lower) than that in the associated background solar wind. The other SFRs are called normal-SFRs. Hence, the SFRs immersed in a certain type of solar wind are categorized into hot-SFRs, cold-SFRs, and normal-SFRs further. The threshold (30$\%$) is higher than the average O$^{7+}$/O$^{6+}$ differences (10$\%$-20$\%$) between the SFRs and associated background solar wind (see the left column in Figure~\ref{fig:1}). 
Different thresholds, such as 50$\%$ and 80$\%$ are also implemented. Quantitively, the proportions and parameters of hot-SFRs (cold-SFRs) are not the same for different thresholds. The higher the threshold, the lower proportions of hot-SFRs (cold-SFRs) in all three types of solar wind. The proportions of hot-SFRs (cold-SFRs) are about 16$\%$ (18$\%$), 9$\%$ (8$\%$), and 4$\%$ (1$\%$) with thresholds of 30$\%$, 50$\%$, and 80$\%$, respectively. The higher the threshold, the more significant the parameter differences between hot-SFRs, cold-SFRs, and normal-SFRs. However, the differences between hot-SFRs, cold-SFRs, and normal-SFRs are qualitatively the same for different thresholds. The numbers and properties of hot-SFRs, cold-SFRs, and normal-SFRS are summarized in \cref{tab:2,tab:3,tab:4}, respectively.

The average duration of the SFRs is the longest (shortest) in AR (CH) wind, with the QS SFRs lying in between (see duration in Table~\ref{tab:1}). 
%The average duration of hot-SFRs and cold-SFRs in CH, AR, and QS solar wind is also the longest (shortest) in AR (CH) wind (see duration in \cref{tab:2,tab:3}). 
%\textbf{The average duration of hot-SFRs and cold-SFRs in CH, AR, and QS solar wind is the shortest in CH wind (see duration in \cref{tab:2,tab:3}). The average duration of cold-SFRs is the longest in AR solar wind while the average duration of hot-SFRs is similar in AR and QS solar wind.}
\textbf{The average duration of hot-SFRs and cold-SFRs in CH, AR, and QS solar wind is generally the longest (shortest) in AR (CH) wind (see duration in \cref{tab:2,tab:3}). The only exception is that the duration of hot-SFRs inside AR and QS solar wind is nearly the same.}
The average duration of hot-SFRs is longer than that of cold-SFRs. The average duration of hot-SFRs (cold-SFRs) inside CH, AR, and QS solar wind is 2.06 $\pm$ 1.39 (1.54 $\pm$ 0.79), 2.47 $\pm$ 1.93 (2.00 $\pm$ 1.45), 2.48 $\pm$ 1.84 (1.84 $\pm$ 1.10) hours, respectively (see duration in \cref{tab:2,tab:3}). 

The proportions of three types of SFRs inside CH, AR, and QS solar wind are analyzed and presented in Figure~\ref{fig:2}(a). The proportions of hot-SFRs (cold-SFRs) inside CH, AR, and QS solar wind are 19$\%$, 15$\%$, and 15$\%$ (22$\%$, 16$\%$, and 18$\%$), respectively. The proportions of hot-SFRs and cold-SFRs are both slightly higher inside CH wind than those in AR and QS solar wind.
The proportion variations with the duration of hot-SFRs, cold-SFRs, and normal-SFRs are presented in Figure~\ref{fig:2}(b). The duration of SFRs is divided into four parts, 1-3 hours, 4-6 hours, 7-9 hours, and 10-12 hours. The statistical results demonstrate that the proportions of hot-SFRs increase with increasing duration (red bars in Figure~\ref{fig:2}(b)). The proportion of hot-SFRs is less than 20$\%$ for the SFRs shorter than 7 hours. While the hot-SFRs dominate about 30$\%$ to 40$\%$ for the SFRs ranging from 7 to 12 hours. Different from hot-SFRs, the proportions of cold-SFRs decrease with increasing duration (blue bars in Figure~\ref{fig:2}(b)). The proportions of cold-SFRs are 19$\%$, 14$\%$, and 11$\%$ with a duration of 0-3 hours, 4-6 hours, and 7-9 hours, respectively. No cold-SFRs last for more than 9 hours.

The magnetic field strengths, proton temperatures, and speeds of hot-SFRs, cold-SFRs, normal-SFRs, and their associated background solar wind are shown in Figure~\ref{fig:3}.
The magnetic field strengths inside hot-SFRs, cold-SFRs, and normal-SFRs are all higher than those in background solar wind (left panels of Figure~\ref{fig:3}). The magnetic field strengths inside cold-SFRs are close to those inside hot-SFRs, which are about 10$\%$ higher than those inside normal-SFRs. The magnetic field strengths of hot-SFRs (cold-SFRs) and normal-SFRs in CH, AR, QS, and solar wind as a whole are 8.74 $\pm$ 3.09 (9.04 $\pm$ 3.07) and 7.89 $\pm$ 2.96, 8.73 $\pm$ 3.46 (8.85 $\pm$ 3.80) and 8.14 $\pm$ 3.22, 9.59 $\pm$ 3.68 (7.50 $\pm$ 2.32) and 7.50 $\pm$ 2.83, and 8.87 $\pm$ 3.39 (8.69 $\pm$ 3.39) and 7.96 $\pm$ 3.09 nT, respectively (see left panels in Figure~\ref{fig:3}).

\begin{table}[]
\centering
\caption{The paramters of cold-SFRs and background in CH, AR, and QS solar wind}
\label{tab:3}
\resizebox{\columnwidth}{!}{%
\begin{threeparttable} 
\begin{tabular}{cccccc}
\hline
 &
   &
  \textbf{MAX} &
  \textbf{DEC} &
  \textbf{MIN} &
  \textbf{TOTAL} \\ \hline
\multirow{4}{*}{\textbf{Number of SFRs}} &
  CH &
  34 &
  80 &
  29 &
  143 \\
 &
  AR &
  97 &
  43 &
  15 &
  155 \\
 &
  QS &
  20 &
  14 &
  28 &
  62 \\
 &
  ALL &
  151 &
  137 &
  72 &
  360 \\ \hline
\multirow{4}{*}{\textbf{Duration (hour) \tnote{a}}} &
  CH &
  1.74 $\pm$ 1.05 &
  1.51 $\pm$ 0.73 &
  1.38 $\pm$ 0.56 &
  1.54 $\pm$ 0.79 \\
 &
  AR &
  2.12 $\pm$ 1.55 &
  1.74 $\pm$ 1.20 &
  1.93 $\pm$ 1.39 &
  2.00 $\pm$ 1.45 \\
 &
  QS &
  1.95 $\pm$ 1.15 &
  1.29 $\pm$ 0.47 &
  2.04 $\pm$ 1.23 &
  1.84 $\pm$ 1.10 \\
 &
  ALL &
  2.01 $\pm$ 1.40 &
  1.56 $\pm$ 0.89 &
  1.75 $\pm$ 1.08 &
  1.79 $\pm$ 1.18 \\ \hline
\multirow{4}{*}{\textbf{O$^{7+}$/O$^{6+}$ \tnote{c}}} &
  CH (SW)\tnote{b} &
  0.07   $\pm$ 0.05 (0.14 $\pm$ 0.12) &
  0.08 $\pm$ 0.07 (0.14 $\pm$ 0.16) &
  0.03 $\pm$ 0.03 (0.06 $\pm$ 0.09) &
  0.07 $\pm$ 0.06 (0.12 $\pm$ 0.14) \\
 &
  AR (SW) &
  0.17 $\pm$ 0.11 (0.28 $\pm$   0.25) &
  0.10 $\pm$ 0.08 (0.19 $\pm$ 0.20) &
  0.05 $\pm$ 0.04 (0.06 $\pm$ 0.07) &
  0.15 $\pm$ 0.11 (0.23 $\pm$ 0.23) \\
 &
  QS (SW) &
  0.09 $\pm$ 0.05 (0.21 $\pm$   0.18) &
  0.11 $\pm$ 0.07 (0.18 $\pm$ 0.15) &
  0.02 $\pm$ 0.01 (0.06 $\pm$ 0.06) &
  0.06 $\pm$ 0.06 (0.13 $\pm$ 0.15) \\
 &
  ALL (SW) &
  0.14 $\pm$ 0.10 (0.24 $\pm$   0.22) &
  0.09 $\pm$ 0.07 (0.16 $\pm$ 0.17) &
  0.03 $\pm$ 0.03 (0.06 $\pm$ 0.07) &
  0.10 $\pm$ 0.10 (0.17 $\pm$ 0.19) \\ \hline
\multirow{4}{*}{\textbf{C$^{6+}$/C$^{5+}$ \tnote{c}}} &
  CH (SW) &
  0.48 $\pm$ 0.30 (0.65 $\pm$   0.42) &
  0.49 $\pm$ 0.35 (0.65 $\pm$ 0.57) &
  0.27 $\pm$ 0.16 (0.42 $\pm$ 0.43) &
  0.45 $\pm$ 0.32 (0.60 $\pm$ 0.52) \\
 &
  AR (SW) &
  0.85 $\pm$ 0.55 (1.13 $\pm$   0.87) &
  0.74 $\pm$ 0.42 (0.96 $\pm$ 0.77) &
  0.47 $\pm$ 0.34 (0.53 $\pm$ 0.54) &
  0.79 $\pm$ 0.52 (1.02 $\pm$ 0.83) \\
 &
  QS (SW) &
  0.58 $\pm$ 0.29 (0.91 $\pm$   0.52) &
  0.77 $\pm$ 0.43 (1.05 $\pm$ 0.68) &
  0.27 $\pm$ 0.16 (0.57 $\pm$ 0.42) &
  0.46 $\pm$ 0.33 (0.78 $\pm$ 0.55) \\
 &
  ALL (SW) &
  0.75 $\pm$ 0.51 (0.99 $\pm$   0.78) &
  0.60 $\pm$ 0.41 (0.79 $\pm$ 0.68) &
  0.32 $\pm$ 0.23 (0.50 $\pm$ 0.46) &
  0.61 $\pm$ 0.46 (0.81 $\pm$ 0.71) \\ \hline
\multirow{4}{*}{\textbf{C$^{6+}$/C$^{4+}$ \tnote{c}}} &
  CH (SW) &
  1.28 $\pm$ 0.99 (2.08 $\pm$   1.72) &
  1.35 $\pm$ 1.30 (2.12 $\pm$ 2.38) &
  0.40 $\pm$ 0.34 (0.95 $\pm$ 1.33) &
  1.16 $\pm$ 1.16 (1.87 $\pm$ 2.10) \\
 &
  AR (SW) &
  2.62 $\pm$ 1.97 (4.28 $\pm$   4.43) &
  2.20 $\pm$ 1.49 (3.69 $\pm$ 5.91) &
  1.08 $\pm$ 0.92 (1.30 $\pm$ 1.68) &
  2.38 $\pm$ 1.84 (3.80 $\pm$ 4.83) \\
 &
  QS (SW) &
  1.78 $\pm$ 1.13 (3.18 $\pm$   2.16) &
  2.24 $\pm$ 1.68 (3.47 $\pm$ 2.93) &
  0.38 $\pm$ 0.28 (1.23 $\pm$ 1.39) &
  1.15 $\pm$ 1.23 (2.34 $\pm$ 2.30) \\
 &
  ALL (SW) &
  2.25 $\pm$ 1.81 (3.62 $\pm$   3.83) &
  1.72 $\pm$ 1.46 (2.80 $\pm$ 4.09) &
  0.54 $\pm$ 0.59 (1.13 $\pm$ 1.44) &
  1.74 $\pm$ 1.65 (2.79 $\pm$ 3.70) \\ \hline
\multirow{4}{*}{\textbf{FIP bias (Fe/O) \tnote{d}}} &
  CH (SW) &
  2.06 $\pm$ 0.71 (2.40 $\pm$   0.90) &
  1.93 $\pm$ 0.59 (2.04 $\pm$ 0.97) &
  1.62 $\pm$ 0.38 (1.96 $\pm$ 1.05) &
  1.91 $\pm$ 0.61 (2.11 $\pm$ 0.99) \\
 &
  AR (SW) &
  2.85 $\pm$ 1.13 (2.86 $\pm$   1.43) &
  2.32 $\pm$ 1.23 (2.39 $\pm$ 1.45) &
  1.62 $\pm$ 0.49 (1.86 $\pm$ 0.75) &
  2.61 $\pm$ 1.18 (2.61 $\pm$ 1.42) \\
 &
  QS (SW) &
  2.12 $\pm$ 0.75 (2.16 $\pm$   0.84) &
  2.22 $\pm$ 0.48 (2.29 $\pm$ 0.76) &
  1.68 $\pm$ 0.43 (1.71 $\pm$ 0.66) &
  1.92 $\pm$ 0.61 (1.98 $\pm$ 0.79) \\
 &
  ALL (SW) &
  2.60 $\pm$ 1.08 (2.66 $\pm$   1.29) &
  2.09 $\pm$ 0.88 (2.18 $\pm$ 1.16) &
  1.65 $\pm$ 0.43 (1.84 $\pm$ 0.86) &
  2.25 $\pm$ 0.99 (2.31 $\pm$ 1.20) \\ \hline
\multirow{4}{*}{\textbf{FIP bias (Mg/O) \tnote{e}}} &
  CH (SW) &
  1.55 $\pm$ 0.33 (1.72 $\pm$   0.44) &
  1.50 $\pm$ 0.29 (1.54 $\pm$ 0.38) &
  1.44 $\pm$ 0.30 (1.43 $\pm$ 0.38) &
  1.50 $\pm$ 0.30 (1.56 $\pm$ 0.41) \\
 &
  AR (SW) &
  2.18 $\pm$ 0.51 (2.30 $\pm$   0.78) &
  1.96 $\pm$ 0.86 (2.05 $\pm$ 0.94) &
  1.48 $\pm$ 0.36 (1.48 $\pm$ 0.43) &
  2.06 $\pm$ 0.63 (2.14 $\pm$ 0.84) \\
 &
  QS (SW) &
  1.57 $\pm$ 0.35 (1.71 $\pm$   0.55) &
  1.69 $\pm$ 0.35 (1.89 $\pm$ 0.61) &
  1.29 $\pm$ 0.20 (1.51 $\pm$ 0.57) &
  1.45 $\pm$ 0.33 (1.65 $\pm$ 0.59) \\
 &
  ALL (SW) &
  1.98 $\pm$ 0.54 (2.09 $\pm$   0.74) &
  1.68 $\pm$ 0.60 (1.75 $\pm$ 0.70) &
  1.38 $\pm$ 0.29 (1.47 $\pm$ 0.47) &
  1.76 $\pm$ 0.57 (1.83 $\pm$ 0.72) \\ \hline
\multirow{4}{*}{\textbf{$A_{He}$ {[}$\%${]}}} &
  CH (SW) &
  3.84 $\pm$ 1.53 (3.48 $\pm$   1.57) &
  3.23 $\pm$ 1.55 (3.19 $\pm$ 1.72) &
  2.22 $\pm$ 1.01 (2.34 $\pm$ 1.26) &
  3.27 $\pm$ 1.56 (3.10 $\pm$ 1.65) \\
 &
  AR (SW) &
  3.23 $\pm$ 2.13 (3.71 $\pm$   2.63) &
  2.42 $\pm$ 1.50 (3.17 $\pm$ 2.53) &
  2.71 $\pm$ 1.25 (2.98 $\pm$ 1.48) &
  2.95 $\pm$ 1.93 (3.46 $\pm$ 2.51) \\
 &
  QS (SW) &
  3.53 $\pm$ 1.20 (3.53 $\pm$   1.21) &
  3.24 $\pm$ 1.19 (3.68 $\pm$ 2.39) &
  2.61 $\pm$ 1.43 (2.67 $\pm$ 1.44) &
  3.01 $\pm$ 1.38 (3.17 $\pm$ 1.68) \\
 &
  ALL (SW) &
  3.41 $\pm$ 1.92 (3.62 $\pm$   2.24) &
  2.97 $\pm$ 1.55 (3.23 $\pm$ 2.08) &
  2.53 $\pm$ 1.31 (2.61 $\pm$ 1.40) &
  3.08 $\pm$ 1.72 (3.26 $\pm$ 2.06) \\ \hline
\multirow{4}{*}{\textbf{\begin{tabular}[c]{@{}c@{}}Magnetic field strength\\ (nT)\end{tabular}}} &
  CH (SW) &
  9.36 $\pm$ 2.60 (7.84 $\pm$   2.67) &
  8.71 $\pm$ 2.77 (8.05 $\pm$ 3.14) &
  9.64 $\pm$ 4.26 (6.44 $\pm$ 3.18) &
  9.04 $\pm$ 3.07 (7.66 $\pm$ 3.12) \\
 &
  AR (SW) &
  8.97 $\pm$ 3.80 (8.33 $\pm$   4.00) &
  9.12 $\pm$ 4.06 (7.65 $\pm$ 4.78) &
  7.36 $\pm$ 2.46 (6.42 $\pm$ 2.47) &
  8.85 $\pm$ 3.80 (7.92 $\pm$ 4.18) \\
 &
  QS (SW) &
  8.27 $\pm$ 2.58 (7.18 $\pm$   2.95) &
  8.24 $\pm$ 2.85 (6.26 $\pm$ 2.43) &
  6.75 $\pm$ 1.59 (5.20 $\pm$ 2.25) &
  7.50 $\pm$ 2.32 (6.01 $\pm$ 2.66) \\
 &
  ALL (SW) &
  8.96 $\pm$ 3.48 (8.07 $\pm$   3.63) &
  8.82 $\pm$ 3.31 (7.72 $\pm$ 3.79) &
  7.78 $\pm$ 3.13 (5.94 $\pm$ 2.76) &
  8.69 $\pm$ 3.39 (7.48 $\pm$ 3.62) \\ \hline
\multirow{4}{*}{\textbf{\begin{tabular}[c]{@{}c@{}}Proton Temperature\\ (10$^{5}$K)\end{tabular}}} &
  CH (SW) &
  1.51 $\pm$ 0.77 (1.28 $\pm$   0.76) &
  1.55 $\pm$ 0.81 (1.46 $\pm$ 0.83) &
  1.61 $\pm$ 0.70 (1.51 $\pm$ 0.82) &
  1.55 $\pm$ 0.78 (1.43 $\pm$ 0.82) \\
 &
  AR (SW) &
  1.22 $\pm$ 0.68 (1.09 $\pm$   0.81) &
  1.46 $\pm$ 1.07 (1.49 $\pm$ 1.18) &
  1.49 $\pm$ 0.76 (1.54 $\pm$ 0.88) &
  1.30 $\pm$ 0.82 (1.26 $\pm$ 0.97) \\
 &
  QS (SW) &
  1.25 $\pm$ 0.50 (1.16 $\pm$   0.53) &
  1.03 $\pm$ 0.45 (1.12 $\pm$ 0.66) &
  1.49 $\pm$ 0.72 (1.01 $\pm$ 0.63) &
  1.33 $\pm$ 0.64 (1.08 $\pm$ 0.61) \\
 &
  ALL (SW) &
  1.28 $\pm$ 0.69 (1.14 $\pm$   0.77) &
  1.47 $\pm$ 0.90 (1.44 $\pm$ 0.97) &
  1.52 $\pm$ 0.73 (1.32 $\pm$ 0.80) &
  1.39 $\pm$ 0.79 (1.30 $\pm$ 0.87) \\ \hline
\multirow{4}{*}{\textbf{\begin{tabular}[c]{@{}c@{}}Proton Number Density\\ (n/cc)\end{tabular}}} &
  CH (SW) &
  9.67 $\pm$ 4.69 (9.30 $\pm$   6.70) &
  7.91 $\pm$ 4.22 (8.42 $\pm$ 6.32) &
  9.52 $\pm$ 2.62 (10.25 $\pm$ 8.61) &
  8.56 $\pm$ 4.32 (8.81 $\pm$ 6.68) \\
 &
  AR (SW) &
  8.13 $\pm$ 8.85 (9.02 $\pm$   8.71) &
  8.42 $\pm$ 12.01 (6.24 $\pm$ 5.94) &
  5.71 $\pm$ 2.32 (7.39 $\pm$ 9.31) &
  8.06 $\pm$ 9.60 (8.11 $\pm$ 8.10) \\
 &
  QS (SW) &
  7.89 $\pm$ 4.31 (8.37 $\pm$   6.44) &
  8.20 $\pm$ 3.36 (6.53 $\pm$ 3.41) &
  6.12 $\pm$ 3.11 (7.09 $\pm$ 5.25) &
  7.47 $\pm$ 3.87 (7.52 $\pm$ 5.48) \\
 &
  ALL (SW) &
  8.40 $\pm$ 7.77 (9.00 $\pm$   8.08) &
  8.12 $\pm$ 7.92 (7.44 $\pm$ 6.05) &
  7.11 $\pm$ 3.21 (8.43 $\pm$ 8.01) &
  8.16 $\pm$ 7.51 (8.28 $\pm$ 7.31) \\ \hline
\multirow{4}{*}{\textbf{\begin{tabular}[c]{@{}c@{}}Speed\\ (km s$^{-1}$)\end{tabular}}} &
  CH (SW) &
  453 $\pm$ 71 (457 $\pm$ 88) &
  493 $\pm$ 86 (505 $\pm$ 99) &
  475 $\pm$ 95 (491 $\pm$ 122) &
  480 $\pm$ 86 (491 $\pm$ 104) \\
 &
  AR (SW) &
  434 $\pm$ 62 (432 $\pm$ 79) &
  475 $\pm$ 89 (484 $\pm$ 104) &
  486 $\pm$ 83 (506 $\pm$ 96) &
  449 $\pm$ 75 (456 $\pm$ 94) \\
 &
  QS (SW) &
  438 $\pm$ 41 (440 $\pm$ 63) &
  421 $\pm$ 49 (434 $\pm$ 69) &
  485 $\pm$ 62 (452 $\pm$ 76) &
  459 $\pm$ 60 (445 $\pm$ 71) \\
 &
  ALL (SW) &
  438 $\pm$ 62 (438 $\pm$ 80) &
  481 $\pm$ 87 (492 $\pm$ 101) &
  482 $\pm$ 78 (479 $\pm$ 103) &
  461 $\pm$ 78 (468 $\pm$ 97) \\ \hline
\multirow{4}{*}{\textbf{\begin{tabular}[c]{@{}c@{}}Expansion Velocity \tnote{f}\\ (km s$^{-1}$)\end{tabular}}} &
  CH &
  -0.03 $\pm$ 9.80 &
  -0.85 $\pm$ 9.64 &
  -0.49 $\pm$ 8.89 &
  -0.58 $\pm$ 9.55 \\
 &
  AR &
  0.50 $\pm$ 9.84 &
  -0.94 $\pm$ 9.57 &
  -0.62 $\pm$ 9.53 &
  -0.06 $\pm$ 9.74 \\
 &
  QS &
  0.26 $\pm$ 9.56 &
  -0.82 $\pm$ 8.67 &
  -2.72 $\pm$ 10.41 &
  -0.94 $\pm$ 9.50 \\
 &
  ALL &
  0.34 $\pm$ 9.78 &
  -0.87 $\pm$ 9.46 &
  -1.30 $\pm$ 9.58 &
  -0.42 $\pm$ 9.63 \\ \hline
\multirow{4}{*}{\textbf{\begin{tabular}[c]{@{}c@{}}Correlation coefficient\\ between speed and O$^{7+}$/O$^{6+}$\end{tabular}}} &
  CH (SW) &
  -0.61 (-0.60) &
  -0.67 (-0.60) &
  -0.59 (-0.67) &
  -0.58 (-0.57) \\
 &
  AR (SW) &
  -0.45 (-0.20) &
  -0.43 (-0.32) &
  -0.82 (-0.74) &
  -0.49 (-0.32) \\
 &
  QS (SW) &
  -0.21 (-0.46) &
  -0.60 (-0.64) &
  -0.54 (-0.55) &
  -0.48 (-0.45) \\
 &
  ALL (SW) &
  -0.45 (-0.29) &
  -0.58 (-0.49) &
  -0.58 (-0.64) &
  -0.51 (-0.41) \\ \hline
 &
   &
   &
   &
   &
  
\end{tabular}%
  \begin{tablenotes}
%\item[ ] \textbf{The values presented in the table are the average and standard derivations of the parameters inside SFRs and background solar wind.}
\item[a] Average duration of SFRs.
\item[b] Background solar wind of SFRs.
\item[c] The O$^{7+}$/O$^{6+}$ (C$^{6+}$/C$^{5+}$, C$^{6+}$/C$^{4+}$) represents the density ratios between O$^{7+}$ and O$^{6+}$ (C$^{6+}$ and C$^{5+}$, C$^{6+}$ and C$^{4+}$), and the values presented in the table are the average and standard derivations of the parameters inside SFRs and background solar wind.
\item[d] The FIP bias is derived from the ratios between Fe/O detected in-situ and their corresponding photospheric values of 0.065 \citep{2009ARA&A..47..481A}.
\item[e] The FIP bias is derived from the ratios between Mg/O detected in-situ and their corresponding photospheric values of 0.081 
\citep {2009ARA&A..47..481A}.
\item[f] The expansion velocity of SFRs is derived from the half of speed difference between the leading and trailing edges of SFRs \citep{ 2016JGRA..121.5005Y,2018ApJS..239...12H}.
   \end{tablenotes}      
    \end{threeparttable} 
}
\end{table}

The proton temperatures demonstrate two characteristics in hot-SFRs, cold-SFRs, normal-SFRs, and their associated background solar wind.
First,the proton temperatures of cold-SFRs are significantly higher than those inside hot-SFRs and normal-SFRs. The proton temperatures of hot-SFRs (cold-SFRs) and normal-SFRs immersed in CH, AR, QS, and solar wind as a whole are 0.94 $\pm$ 0.60 (1.55 $\pm$ 0.78) and 1.07 $\pm$ 0.66, 0.76 $\pm$ 0.63 (1.30 $\pm$ 0.82) and 0.84 $\pm$ 0.70, 0.78 $\pm$ 0.46 (1.33 $\pm$ 0.64) and 0.84 $\pm$ 0.55, and 0.83 $\pm$ 0.60 (1.39 $\pm$ 0.79) and 0.91 $\pm$ 0.67 $\times$ 10$^{5}$K, respectively.
Second, the proton temperatures of cold-SFRs are almost the same as or slightly higher than those of background solar wind, while the proton temperatures of hot-SFRs and normal-SFRs are clearly lower {as} compared with background solar wind. The proton temperatures inside hot-SFRs (normal-SFRs) are about 30$\%$ (20$\%$) lower than those inside the associated background solar wind (red and green lines in the middle panels of Figure~\ref{fig:3}).

The significant characteristic in the right panels of Figure~\ref{fig:3} is that the speeds of cold-SFRs are all higher than those of hot-SFRs and normal-SFRs in all three types of solar wind. The speeds of hot-SFRs and normal-SFRs are nearly the same. The speeds of hot-SFRs (cold-SFRs) and normal-SFRs in CH, AR, QS, and solar wind as a whole are 444 $\pm$ 89 (480 $\pm$ 86) and 462 $\pm$ 90, 417 $\pm$ 66 (449 $\pm$ 75) and 412 $\pm$ 66, 429 $\pm$ 63 (459 $\pm$ 60) and 420 $\pm$ 64, and 428 $\pm$ 75 and (461 $\pm$ 78) and 426 $\pm$ 76 km s$^{-1}$, respectively. Statistically, the hot-SFRs, cold-SFRs, and normal-SFRs immersed in three types of solar wind all not expand significantly. The expansion velocities are all nearly zero (see expansion velocities in \cref{tab:2,tab:3,tab:4}).

The compositional signatures of hot-SFRs, cold-SFRs, normal-SFRs, and associated background solar wind are shown in Figure~\ref{fig:4}.
The O$^{7+}$/O$^{6+}$ is shown in the left panels. There are two significant characteristics of the O$^{7+}$/O$^{6+}$ inside hot-SFRs, cold-SFRs, normal-SFRs, and their background solar wind.
First, the O$^{7+}$/O$^{6+}$ inside hot-SFRs (cold-SFRs) is significantly higher (lower) than that inside normal-SFRs. The O$^{7+}$/O$^{6+}$ inside hot-SFRs (cold-SFRs) is about 60$\%$ higher (lower) than that inside normal-SFRs. The O$^{7+}$/O$^{6+}$ inside hot-SFRs (cold-SFRs) and normal-SFRs immersed in CH, AR, QS, and the solar wind as a whole is 0.24 $\pm$ 0.20 (0.07 $\pm$ 0.06) and 0.14 $\pm$ 0.10, 0.43 $\pm$ 0.27 (0.15 $\pm$ 0.11) and 0.26 $\pm$ 0.15, 0.32 $\pm$ 0.31 (0.06 $\pm$ 0.06) and 0.20 $\pm$ 0.16, and 0.35 $\pm$ 0.27 (0.10 $\pm$ 0.10) and 0.22 $\pm$ 0.15, respectively.
Second, the O$^{7+}$/O$^{6+}$ inside hot-SFRs (cold-SFRs) is significantly higher (lower) than that inside the background solar wind (red and blue lines in the left panels of Figure~\ref{fig:4}). 
Meanwhile, the O$^{7+}$/O$^{6+}$ inside normal-SFRs is about 10$\%$-20$\%$ higher than that inside associated background solar wind (green lines in the left panels of Figure~\ref{fig:4}). The O$^{7+}$/O$^{6+}$ of hot-SFRs (cold-SFRs) is about 100$\%$ higher (50$\%$ lower) than that of background solar wind. The discrepancy between the O$^{7+}$/O$^{6+}$ inside hot-SFRs (cold-SFRs) and background solar wind is well above the threshold (30$\%$) for selecting hot-SFRs (cold-SFRs).
The characteristics of $A_{He}$ are presented in the middle column of Figure~\ref{fig:4}. Generally, the $A_{He}$ inside hot-SFRs (cold-SFRs) is the highest (lowest), with the $A_{He}$ inside normal-SFRs lying in between. The $A_{He}$ inside hot-SFRs (cold-SFRs) is about 10$\%$-20$\%$ higher (lower) than that in the background solar wind. The $A_{He}$ is almost the same inside normal-SFRs and in their background solar wind (green lines in middle panels of Figure~\ref{fig:4}). The $A_{He}$ distribution ranges of hot-SFRs (cold-SFRs) are also wider (narrower) than those of normal-SFRs (see shaded areas in the middle panels of Figure~\ref{fig:4}).
The FIP bias presented in Figure~\ref{fig:4} is derived from Fe/O. The statistical results demonstrate that the FIP bias (distribution ranges) of cold-SFRs are all lower (narrower) than those of hot-SFRs and normal-SFRs in all three types of solar wind (right panels in Figure~\ref{fig:4}). The FIP bias inside hot-SFRs (cold-SFRs) and normal-SFRs immersed in CH, AR, QS, and solar wind as a whole is 2.37 $\pm$ 1.37 (1.91 $\pm$ 0.61) and 2.30 $\pm$ 1.01, 3.13 $\pm$ 1.83 (2.61 $\pm$ 1.18) and 3.05 $\pm$ 1.54, 2.35 $\pm$ 0.96 (1.92 $\pm$ 0.61) and 2.59 $\pm$ 1.25, and 2.73 $\pm$ 1.60 (2.25 $\pm$ 0.99) and 2.76 $\pm$ 1.40, respectively. The distribution ranges for hot-SFRs (cold-SFRs) inside CH, AR, and QS solar wind are about 1.0-3.8 (1.3-2.5), 1.3-5.0 (1.5-4.8), and 1.4-3.3 (1.3-2.5), respectively. The characteristics of FIP bias derived from Mg/O are nearly the same as those deduced from Fe/O (see \cref{tab:1,tab:2,tab:3,tab:4}).

\begin{table}[]
\centering
\caption{The paramters of normal-SFRs and background in CH, AR, and QS solar wind}
\label{tab:4}
\resizebox{\columnwidth}{!}{%
\begin{threeparttable} 
\begin{tabular}{cccccc}
\hline
 &
   &
  \textbf{MAX} &
  \textbf{DEC} &
  \textbf{MIN} &
  \textbf{TOTAL} \\ \hline
\multirow{4}{*}{\textbf{Number of SFRs}} &
  CH &
  141 &
  205 &
  40 &
  386 \\
 &
  AR &
  486 &
  180 &
  21 &
  687 \\
 &
  QS &
  118 &
  96 &
  25 &
  239 \\
 &
  ALL &
  745 &
  481 &
  86 &
  1312 \\ \hline
\multirow{4}{*}{\textbf{Duration (hour) \tnote{a}}} &
  CH &
  1.88 $\pm$ 1.20 &
  1.91 $\pm$ 1.09 &
  2.05 $\pm$ 1.66 &
  1.91 $\pm$ 1.20 \\
 &
  AR &
  2.23 $\pm$ 1.56 &
  2.18 $\pm$ 1.51 &
  1.43 $\pm$ 0.81 &
  2.20 $\pm$ 1.53 \\
 &
  QS &
  2.05 $\pm$ 1.30 &
  1.99 $\pm$ 1.29 &
  2.44 $\pm$ 2.45 &
  2.07 $\pm$ 1.45 \\
 &
  ALL &
  2.14 $\pm$ 1.46 &
  2.02 $\pm$ 1.31 &
  2.01 $\pm$ 1.81 &
  2.09 $\pm$ 1.43 \\ \hline
\multirow{4}{*}{\textbf{O$^{7+}$/O$^{6+}$ \tnote{c}}} &
  CH (SW) \tnote{b} &
  0.16 $\pm$   0.10 (0.15 $\pm$ 0.12) &
  0.14 $\pm$ 0.10 (0.11 $\pm$ 0.12) &
  0.08 $\pm$ 0.06 (0.06 $\pm$ 0.08) &
  0.14 $\pm$ 0.10 (0.12 $\pm$ 0.11) \\
 &
  AR (SW) &
  0.27 $\pm$ 0.16 (0.25 $\pm$   0.19) &
  0.22 $\pm$ 0.11 (0.19 $\pm$ 0.14) &
  0.12 $\pm$ 0.05 (0.12 $\pm$ 0.08) &
  0.26 $\pm$ 0.15 (0.23 $\pm$ 0.18) \\
 &
  QS (SW) &
  0.25 $\pm$ 0.20 (0.20 $\pm$   0.19) &
  0.19 $\pm$ 0.09 (0.17 $\pm$ 0.13) &
  0.06 $\pm$ 0.04 (0.05 $\pm$ 0.05) &
  0.20 $\pm$ 0.16 (0.17 $\pm$ 0.16) \\
 &
  ALL (SW) &
  0.25 $\pm$ 0.17 (0.22 $\pm$   0.19) &
  0.18 $\pm$ 0.11 (0.15 $\pm$ 0.13) &
  0.08 $\pm$ 0.06 (0.07 $\pm$ 0.08) &
  0.22 $\pm$ 0.15 (0.18 $\pm$ 0.17) \\ \hline
\multirow{4}{*}{\textbf{C$^{6+}$/C$^{5+}$ \tnote{c}}} &
  CH (SW) &
  0.87 $\pm$ 0.43 (0.81 $\pm$   0.51) &
  0.74 $\pm$ 0.48 (0.65 $\pm$ 0.53) &
  0.54 $\pm$ 0.34 (0.45 $\pm$ 0.41) &
  0.76 $\pm$ 0.46 (0.68 $\pm$ 0.52) \\
 &
  AR (SW) &
  1.21 $\pm$ 0.76 (1.11 $\pm$   0.74) &
  1.24 $\pm$ 0.67 (1.04 $\pm$ 0.63) &
  0.98 $\pm$ 0.41 (0.83 $\pm$ 0.51) &
  1.21 $\pm$ 0.73 (1.08 $\pm$ 0.71) \\
 &
  QS (SW) &
  1.15 $\pm$ 0.65 (1.08 $\pm$   0.84) &
  1.12 $\pm$ 0.51 (1.01 $\pm$ 0.62) &
  0.55 $\pm$ 0.29 (0.51 $\pm$ 0.37) &
  1.06 $\pm$ 0.59 (0.96 $\pm$ 0.72) \\
 &
  ALL (SW) &
  1.14 $\pm$ 0.71 (1.04 $\pm$   0.73) &
  1.01 $\pm$ 0.61 (0.85 $\pm$ 0.62) &
  0.62 $\pm$ 0.37 (0.54 $\pm$ 0.45) &
  1.06 $\pm$ 0.67 (0.92 $\pm$ 0.69) \\ \hline
\multirow{4}{*}{\textbf{C$^{6+}$/C$^{4+}$ \tnote{c}}} &
  CH (SW) &
  2.99 $\pm$ 2.00 (2.74 $\pm$   2.23) &
  2.47 $\pm$ 2.16 (2.18 $\pm$ 2.46) &
  1.19 $\pm$ 0.96 (1.08 $\pm$ 1.36) &
  2.51 $\pm$ 2.07 (2.25 $\pm$ 2.33) \\
 &
  AR (SW) &
  4.49 $\pm$ 3.63 (4.03 $\pm$   3.73) &
  4.63 $\pm$ 4.73 (3.65 $\pm$ 3.19) &
  2.38 $\pm$ 1.08 (2.38 $\pm$ 1.86) &
  4.49 $\pm$ 3.92 (3.88 $\pm$ 3.56) \\
 &
  QS (SW) &
  4.26 $\pm$ 3.32 (3.74 $\pm$   3.08) &
  3.89 $\pm$ 2.13 (3.40 $\pm$ 2.95) &
  1.20 $\pm$ 0.90 (1.13 $\pm$ 1.34) &
  3.74 $\pm$ 2.86 (3.21 $\pm$ 2.96) \\
 &
  ALL (SW) &
  4.21 $\pm$ 3.41 (3.71 $\pm$   3.42) &
  3.61 $\pm$ 3.55 (2.89 $\pm$ 2.95) &
  1.40 $\pm$ 1.05 (1.35 $\pm$ 1.55) &
  3.82 $\pm$ 3.43 (3.20 $\pm$ 3.20) \\ \hline
\multirow{4}{*}{\textbf{FIP bias (Fe/O) \tnote{d}}} &
  CH (SW) &
  2.50 $\pm$ 0.98 (2.31 $\pm$   0.92) &
  2.24 $\pm$ 1.06 (2.02 $\pm$ 0.82) &
  1.96 $\pm$ 0.73 (1.84 $\pm$ 0.72) &
  2.30 $\pm$ 1.01 (2.10 $\pm$ 0.86) \\
 &
  AR (SW) &
  3.09 $\pm$ 1.61 (2.89 $\pm$   1.45) &
  2.93 $\pm$ 1.35 (2.58 $\pm$ 1.25) &
  2.81 $\pm$ 1.10 (2.34 $\pm$ 0.97) &
  3.05 $\pm$ 1.54 (2.79 $\pm$ 1.40) \\
 &
  QS (SW) &
  2.86 $\pm$ 1.57 (2.50 $\pm$   1.05) &
  2.41 $\pm$ 0.77 (2.17 $\pm$ 0.75) &
  2.06 $\pm$ 0.67 (1.79 $\pm$ 0.90) &
  2.59 $\pm$ 1.25 (2.26 $\pm$ 0.95) \\
 &
  ALL (SW) &
  2.96 $\pm$ 1.53 (2.71 $\pm$   1.33) &
  2.55 $\pm$ 1.18 (2.24 $\pm$ 1.02) &
  2.14 $\pm$ 0.84 (1.91 $\pm$ 0.87) &
  2.76 $\pm$ 1.40 (2.46 $\pm$ 1.22) \\ \hline
\multirow{4}{*}{\textbf{FIP bias (Mg/O) \tnote{e}}} &
  CH (SW) &
  1.78 $\pm$ 0.43 (1.74 $\pm$   0.48) &
  1.59 $\pm$ 0.42 (1.57 $\pm$ 0.47) &
  1.44 $\pm$ 0.25 (1.44 $\pm$ 0.39) &
  1.64 $\pm$ 0.43 (1.61 $\pm$ 0.48) \\
 &
  AR (SW) &
  2.26 $\pm$ 0.63 (2.21 $\pm$   0.66) &
  2.23 $\pm$ 0.86 (2.04 $\pm$ 0.75) &
  1.86 $\pm$ 0.53 (1.90 $\pm$ 0.48) &
  2.25 $\pm$ 0.70 (2.15 $\pm$ 0.69) \\
 &
  QS (SW) &
  2.02 $\pm$ 0.65 (1.85 $\pm$   0.55) &
  1.88 $\pm$ 0.60 (1.74 $\pm$ 0.46) &
  1.55 $\pm$ 0.38 (1.52 $\pm$ 0.56) &
  1.91 $\pm$ 0.62 (1.76 $\pm$ 0.53) \\
 &
  ALL (SW) &
  2.14 $\pm$ 0.63 (2.06 $\pm$   0.65) &
  1.90 $\pm$ 0.72 (1.76 $\pm$ 0.62) &
  1.55 $\pm$ 0.39 (1.55 $\pm$ 0.50) &
  2.02 $\pm$ 0.67 (1.90 $\pm$ 0.65) \\ \hline
\multirow{4}{*}{\textbf{$A_{He}$ {[}$\%${]}}} &
  CH (SW) &
  3.40 $\pm$ 1.57 (3.43 $\pm$   1.52) &
  3.13 $\pm$ 1.61 (3.11 $\pm$ 1.56) &
  2.74 $\pm$ 1.35 (2.53 $\pm$ 1.21) &
  3.18 $\pm$ 1.58 (3.16 $\pm$ 1.54) \\
 &
  AR (SW) &
  3.67 $\pm$ 2.58 (3.46 $\pm$   2.33) &
  3.01 $\pm$ 1.86 (3.30 $\pm$ 2.29) &
  2.00 $\pm$ 1.19 (2.30 $\pm$ 1.22) &
  3.47 $\pm$ 2.42 (3.38 $\pm$ 2.30) \\
 &
  QS (SW) &
  3.62 $\pm$ 1.97 (3.43 $\pm$   1.91) &
  3.34 $\pm$ 2.11 (3.34 $\pm$ 1.90) &
  1.84 $\pm$ 1.36 (2.21 $\pm$ 1.19) &
  3.36 $\pm$ 2.04 (3.23 $\pm$ 1.87) \\
 &
  ALL (SW) &
  3.61 $\pm$ 2.34 (3.44 $\pm$   2.11) &
  3.12 $\pm$ 1.81 (3.21 $\pm$ 1.89) &
  2.36 $\pm$ 1.38 (2.40 $\pm$ 1.21) &
  3.37 $\pm$ 2.15 (3.27 $\pm$ 1.99) \\ \hline
\multirow{4}{*}{\textbf{\begin{tabular}[c]{@{}c@{}}Magnetic field strength\\ (nT)\end{tabular}}} &
  CH (SW) &
  7.79 $\pm$ 2.78 (7.33 $\pm$   3.21) &
  8.11 $\pm$ 3.17 (7.00 $\pm$ 2.89) &
  7.18 $\pm$ 2.27 (5.78 $\pm$ 2.42) &
  7.89 $\pm$ 2.96 (6.96 $\pm$ 2.99) \\
 &
  AR (SW) &
  8.21 $\pm$ 3.24 (7.25 $\pm$   3.15) &
  8.05 $\pm$ 3.27 (6.94 $\pm$ 3.12) &
  6.91 $\pm$ 1.43 (5.18 $\pm$ 1.60) &
  8.14 $\pm$ 3.22 (7.10 $\pm$ 3.13) \\
 &
  QS (SW) &
  7.98 $\pm$ 3.17 (7.15 $\pm$   2.90) &
  7.02 $\pm$ 2.42 (6.04 $\pm$ 2.29) &
  7.04 $\pm$ 2.15 (5.22 $\pm$ 2.08) &
  7.50 $\pm$ 2.83 (6.40 $\pm$ 2.65) \\
 &
  ALL (SW) &
  8.10 $\pm$ 3.16 (7.24 $\pm$   3.14) &
  7.87 $\pm$ 3.11 (6.80 $\pm$ 2.90) &
  7.08 $\pm$ 2.09 (5.47 $\pm$ 2.21) &
  7.96 $\pm$ 3.09 (6.92 $\pm$ 3.02) \\ \hline
\multirow{4}{*}{\textbf{\begin{tabular}[c]{@{}c@{}}Proton Temperature\\ (10$^{5}$K)\end{tabular}}} &
  CH (SW) &
  1.02 $\pm$ 0.68 (1.22 $\pm$   0.77) &
  1.07 $\pm$ 0.63 (1.44 $\pm$ 0.82) &
  1.24 $\pm$ 0.67 (1.55 $\pm$ 0.80) &
  1.07 $\pm$ 0.66 (1.38 $\pm$ 0.81) \\
 &
  AR (SW) &
  0.81 $\pm$ 0.57 (1.05 $\pm$   0.77) &
  0.94 $\pm$ 0.99 (1.16 $\pm$ 0.86) &
  0.79 $\pm$ 0.54 (1.00 $\pm$ 0.62) &
  0.84 $\pm$ 0.70 (1.08 $\pm$ 0.79) \\
 &
  QS (SW) &
  0.88 $\pm$ 0.61 (1.01 $\pm$   0.61) &
  0.80 $\pm$ 0.47 (1.09 $\pm$ 0.64) &
  0.86 $\pm$ 0.52 (1.14 $\pm$ 0.61) &
  0.84 $\pm$ 0.55 (1.06 $\pm$ 0.62) \\
 &
  ALL (SW) &
  0.86 $\pm$ 0.60 (1.08 $\pm$   0.76) &
  0.97 $\pm$ 0.77 (1.29 $\pm$ 0.82) &
  1.03 $\pm$ 0.64 (1.28 $\pm$ 0.74) &
  0.91 $\pm$ 0.67 (1.18 $\pm$ 0.79) \\ \hline
\multirow{4}{*}{\textbf{\begin{tabular}[c]{@{}c@{}}Proton Number Density\\ (n/cc)\end{tabular}}} &
  CH (SW) &
  8.56 $\pm$ 6.41 (8.13 $\pm$   6.99) &
  8.58 $\pm$ 8.10 (6.73 $\pm$ 5.33) &
  9.79 $\pm$ 9.22 (7.11 $\pm$ 4.85) &
  8.65 $\pm$ 7.59 (7.28 $\pm$ 6.02) \\
 &
  AR (SW) &
  8.17 $\pm$ 7.04 (7.47 $\pm$   6.86) &
  7.59 $\pm$ 5.53 (6.24 $\pm$ 5.40) &
  8.44 $\pm$ 4.26 (6.92 $\pm$ 4.27) &
  8.03 $\pm$ 6.67 (7.14 $\pm$ 6.50) \\
 &
  QS (SW) &
  8.23 $\pm$ 7.64 (7.79 $\pm$   6.78) &
  8.03 $\pm$ 6.14 (6.72 $\pm$ 4.79) &
  20.81 $\pm$ 12.75 (7.40 $\pm$ 5.29) &
  8.55 $\pm$ 7.60 (7.32 $\pm$ 6.00) \\
 &
  ALL (SW) &
  8.24 $\pm$ 7.03 (7.53 $\pm$   6.70) &
  8.09 $\pm$ 6.84 (6.57 $\pm$ 5.32) &
  11.20 $\pm$ 9.87 (7.01 $\pm$ 4.79) &
  8.28 $\pm$ 7.10 (7.14 $\pm$ 6.16) \\ \hline
\multirow{4}{*}{\textbf{\begin{tabular}[c]{@{}c@{}}Speed\\ (km s$^{-1}$)\end{tabular}}} &
  CH (SW) &
  440 $\pm$ 85 (465 $\pm$ 101) &
  475 $\pm$ 92 (518 $\pm$ 107) &
  468 $\pm$ 84 (496 $\pm$ 112) &
  462 $\pm$ 90 (498 $\pm$ 108) \\
 &
  AR (SW) &
  406 $\pm$ 58 (426 $\pm$ 77) &
  427 $\pm$ 82 (450 $\pm$ 94) &
  416 $\pm$ 65 (449 $\pm$ 92) &
  412 $\pm$ 66 (433 $\pm$ 83) \\
 &
  QS (SW) &
  418 $\pm$ 66 (439 $\pm$ 75) &
  423 $\pm$ 61 (449 $\pm$ 84) &
  419 $\pm$ 64 (459 $\pm$ 89) &
  420 $\pm$ 64 (446 $\pm$ 82) \\
 &
  ALL (SW) &
  413 $\pm$ 66 (437 $\pm$ 85) &
  446 $\pm$ 86 (483 $\pm$ 105) &
  441 $\pm$ 78 (476 $\pm$ 103) &
  426 $\pm$ 76 (458 $\pm$ 97) \\ \hline
\multirow{4}{*}{\textbf{\begin{tabular}[c]{@{}c@{}}Expansion Velocity \tnote{f}\\ (km s$^{-1}$)\end{tabular}}} &
  CH &
  0.69 $\pm$ 9.20 &
  -0.64 $\pm$ 10.07 &
  -0.98 $\pm$ 9.15 &
  -0.43 $\pm$ 9.73 \\
 &
  AR &
  0.09 $\pm$ 9.32 &
  -0.40 $\pm$ 8.76 &
  -0.01 $\pm$ 8.09 &
  -0.11 $\pm$ 9.03 \\
 &
  QS &
  -0.21 $\pm$ 9.03 &
  -1.00 $\pm$ 8.41 &
  -3.27 $\pm$ 9.65 &
  -1.30 $\pm$ 8.95 \\
 &
  ALL &
  0.17 $\pm$ 9.24 &
  -0.61 $\pm$ 9.33 &
  -1.51 $\pm$ 9.17 &
  -0.45 $\pm$ 9.28 \\ \hline
\multirow{4}{*}{\textbf{\begin{tabular}[c]{@{}c@{}}Correlation coefficient\\ between speed and O$^{7+}$/O$^{6+}$\end{tabular}}} &
  CH (SW) &
  -0.58 (-0.62) &
  -0.64 (-0.63) &
  -0.62 (-0.70) &
  -0.61 (-0.63) \\
 &
  AR (SW) &
  -0.41 (-0.43) &
  -0.35 (-0.47) &
  -0.68 (-0.65) &
  -0.39 (-0.44) \\
 &
  QS (SW) &
  -0.20 (-0.19) &
  -0.27 (-0.44) &
  -0.55 (-0.60) &
  -0.21 (-0.30) \\
 &
  ALL (SW) &
  -0.40 (-0.43) &
  -0.51 (-0.57) &
  -0.57 (-0.66) &
  -0.44 (-0.49) \\ \hline  
 &
   &
   &
   &
   &
\end{tabular}%
  \begin{tablenotes}
%\item[ ] \textbf{The values presented in the table are the average and standard derivations of the parameters inside SFRs and background solar wind.}
\item[a] Average duration of SFRs.
\item[b] Background solar wind of SFRs.
\item[c] The O$^{7+}$/O$^{6+}$ (C$^{6+}$/C$^{5+}$, C$^{6+}$/C$^{4+}$) represents the density ratios between O$^{7+}$ and O$^{6+}$ (C$^{6+}$ and C$^{5+}$, C$^{6+}$ and C$^{4+}$), and the values presented in the table are the average and standard derivations of the parameters inside SFRs and background solar wind.
\item[d] The FIP bias is derived from the ratios between Fe/O detected in-situ and their corresponding photospheric values of 0.065 \citep{2009ARA&A..47..481A}.
\item[e] The FIP bias is derived from the ratios between Mg/O detected in-situ and their corresponding photospheric values of 0.081 
\citep {2009ARA&A..47..481A}.
\item[f] The expansion velocity of SFRs is derived from the half of speed difference between the leading and trailing edges of SFRs \citep{ 2016JGRA..121.5005Y,2018ApJS..239...12H}.
   \end{tablenotes}      
    \end{threeparttable} 
}
\end{table}

The distribution characteristics in speed versus O$^{7+}$/O$^{6+}$ space of hot-SFRs, cold-SFRs, normal-SFRs, and associated background solar wind are also examined. The distributions for normal-SFRs, hot-SFRs, and cold-SFRs immersed in CH, AR, and QS solar wind are presented in the left, middle, and right columns of Figure~\ref{fig:5}. There is a significant anti-correlation between the speed and O$^{7+}$/O$^{6+}$ in the background solar wind. 
The CCs and linear fitting results of normal-SFRs are similar to those of associated background solar wind (left panels in Figure~\ref{fig:5}). The CCs of normal-SFRs (background solar wind) immersed in CH, AR, QS, and solar wind as a whole are -0.61 (-0.63), -0.39 (-0.44), -0.21 (-0.30), and -0.44 (-0.49), respectively.
The slopes and intercepts of normal-SFRs (background solar wind) are -0.0030 (-0.0031) and 0.39 (0.40), -0.0021 (-0.0024) and 0.20 (0.30), -0.0019 (-0.0026) and -0.01 (0.22), and -0.0027 (-0.0030) and 0.39 (0.47), respectively.
The distribution characteristics of hot-SFRs are different from the associated background solar wind. The CCs of hot-SFRs are lower than those of normal-SFRs and background solar wind. In addition, the slopes and intercepts of hot-SFRs are also different from those of normal-SFRs and background solar wind. The CCs of hot-SFRs (background solar wind) between the speed and O$^{7+}$/O$^{6+}$ immersed in CH, AR, QS, and solar wind as a whole are -0.46 (-0.54), -0.29 (-0.42), -0.12 (-0.35), and -0.34 (-0.47), respectively. The slopes and intercepts of hot-SFRs (background solar wind) are, -0.0028 (-0.0031) and 0.50 (0.41), -0.0019 (-0.0028) and 0.33 (0.42), -0.0011 (-0.0031) and -0.17 (0.46), and -0.0025 (-0.0032) and 0.47 (0.54), respectively.
The CCs and slops of cold-SFRs are nearly the same as those of associated background solar wind (right panels in Figure~\ref{fig:5}). The difference is that the intercepts of cold-SFRs are all lower than those of background in all three types of solar wind. This is consistent with the fact that the O$^{7+}$/O$^{6+}$ inside cold-SFRs is lower than that of background solar wind.
The CCs of cold-SFRs (background solar wind) immersed in CH, AR, QS, and solar wind as a whole are -0.58 (-0.57), -0.49 (-0.32), -0.48 (-0.45), and -0.51 (-0.41), respectively.
The slopes and intercepts of cold-SFRs (background solar wind) immersed in CH, AR, QS, and solar wind as a whole are -0.0034 (-0.0038) and 0.29 (0.71), -0.0028 (-0.0027) and 0.29 (0.38), -0.0039 (-0.0039) and 0.41 (0.64), -0.0035 (-0.0034) and 0.45 (0.60), respectively.

The anti-correlation between the solar wind speed and the O$^{7+}$/O$^{6+}$ was first discovered from the measurements of the International Sun-Earth Explorer \citep{1989SoPh..124..167O} and then confirmed by the detection of Ulysses \citep{1995SSRv...72...49G,2003JGRA..108.1158G}.
In addition, the anti-correlation is still valid in the solar wind originating from different source regions, and the slopes and intercepts are almost the same for the solar wind coming from CH, AR, and QS regions \citep{2017ApJ...836..169F}. There are two main solar wind models, the reconnection-loop-opening (RLO) models \citep{1999JGR...10419765F,2003JGRA..108.1157F,2004ApJ...612.1171W} and wave-turbulence-driven (WTD) models \citep{1986JGR....91.4111H,1991ApJ...372L..45W,2007ApJS..171..520C}, for describing the releasing, heating, and acceleration of solar wind. The previous statistical results indicated that the two mechanisms are both valid \citep{2017ApJ...836..169F,2018MNRAS.478.1884F}. The anti-correlation between the solar wind speed and the O$^{7+}$/O$^{6+}$ can be reasonably interpreted by the above two main solar wind generation mechanisms \citep{2017ApJ...836..169F}.
The present results demonstrate that the anti-correlation between the speed and the O$^{7+}$/O$^{6+}$ is still valid for the plasma inside SFRs, especially for normal-SFRs and cold-SFRs. The CCs, slopes, and intercepts of plasma inside normal-SFRs are similar to those of background solar wind. Therefore, we suggest that the generation mechanisms of plasma inside normal-SFRs are the same as those of their background in all three types of solar wind. In contrast, the anti-correlation between speed and O$^{7+}$/O$^{6+}$ is not so significant for hot-SFRs. The slopes and intercepts of hot-SFRs are different from those of background solar wind. The above results indicate that the plasma generation mechanism of hot-SFRs may not be totally the same as that of background solar wind. The CCs and slopes (intercepts) of cold-SFRs are similar to (lower than) those of background solar wind, which means that the cold-SFRs should also originate from the Sun and be associated with lower {electron} temperatures in source regions.

The present statistical results demonstrate that the properties of hot-SFRs and cold-SFRs are significantly different. They seemed to be lying in the two extremes {as} compared with normal-SFRs and background solar wind.
Statistically, the hot-SFRs are associated with longer duration, lower speeds and proton temperatures, higher charge states and helium abundance as compared with normal-SFRs. The cold-SFRs are generally accompanied by shorter duration, higher speeds and proton temperatures, lower charge states, helium abundance, and FIP bias as compared with normal-SFRs and background solar wind. The same characteristic of hot-SFRs and cold-SFRs is that the magnetic field strengths are both higher than those of normal-SFRs. 
The hot-SFRs and cold-SFRs should come from the Sun. As the charge states and element abundances should be almost the same inside SFRs and background solar wind if SFRs are generated in the heliosphere. It should point out, however, the same properties inside SFRs and background solar wind do not only mean that the SFRs are generated in the heliosphere.

The {electron} temperature inside hot-SFRs ought to be higher than the background in the source regions as the O$^{7+}$/O$^{6+}$ ratios represent the temperature near the Sun.
There are several possible originations of the hot-SFRs. 
First, the hot-SFRs come from the streamers associated with plasma blobs. The remote and in-situ joint observations demonstrate that the SFRs can also be released at the tips of streamers \citep{2009SoPh..256..307R, 2010JGRA..115.4103R, 2010JGRA..115.4104R,2011ApJ...734....7R, 2017ApJ...835L...7S, 2019ApJ...882...51S}. The above studies provide direct evidence for the notion that the SFRs come from the Sun. 
%The SFRs coming from the streamers should be associated with higher O$^{7+}$/O$^{6+}$ 
The O$^{7+}$/O$^{6+}$ of SFRs coming from streamers should be similar to or higher than those of background solar wind {as the temperature of streamers is higher than surrounding regions \citep{2010ApJ...708.1650H,2018ApJ...859..155B}.} In addition, the SFR release processes in streamers are generally associated with magnetic reconnection. The plasma inside SFRs could be heated during magnetic reconnection. Thus the O$^{7+}$/O$^{6+}$ ratios should be higher inside SFRs released from streamers than those of background solar wind, statistically.
Second, the hot-SFRs may be produced by small-scale activities on the Sun. The statistical properties of hot-SFRs can be explained reasonably by the scenario. It is known that the plasmas with higher charge states inside ICMEs are heated during the flaring processes \citep{2004JGRA..109.1112L, 2013SoPh..284...17G, 2020ApJ...900L..18F,2022ApJ...928..136Z}. The heating processes for hot-SFRs may be the same as the CMEs. Therefore, the plasma inside hot-SFRs may be heated by the small-scale eruption processes. In addition, the statistical properties of hot-SFRs are similar to the ICMEs, qualitatively. Hence, some hot-SFRs may be generated by small-scale eruptions on the Sun. 
Third, the generation of hot-SFRs can also be connected with magnetic erosion processes in the heliosphere. Previous studies demonstrated that magnetic reconnection can occur at the front and rear of MCs \citep{2006A&A...455..349D, 2007SoPh..244..115D,2012JGRA..117.9101R,2014JGRA..119.7088J,2014JGRA..119...26L} and SFRs \citep{2010ApJ...720..454T}. The magnetic reconnection can erode parts of magnetic field and hence reduce the scale of magnetic flux ropes. Therefore, the SFRs can also be generated by magnetic erosion processes during propagation in the heliosphere. In this scenario, the properties of SFRs should be similar to those of MCs. {Moreover, the A$_{He}$ inside hot-SFRs is overall higher than that inside background, implying the small-scale solar activities and magnetic erosion processes may contribute more to the formation of hot-SFRs. Considering that streamer belt solar wind generally shows low A$_{He}$ \citep{2007ApJ...660..901K,2012ApJ...745..162K, 2009JGRA..114.4103S,2018MNRAS.478.1884F}.}

What is the generation mechanism of the cold-SFRs? One of the reasonable interpretations is that the cold-SFRs are associated with small cold filament or prominence eruptions. Generally, the materials of filaments are associated with lower charge states as they are cold on the Sun \citep{2010ApJ...723L..22L}. In addition, the FIP bias of filaments should be lower {as} compared with corona, considering that the materials of filaments come from the chromosphere \citep{2017ApJ...836L..11S}. Qualitatively, the properties of cold-SFRs are consistent with the above two characteristics (lower charge states and FIP bias). 
However, the helium abundance of cold-SFRs may be not consistent with that of filaments. Generally, it is believed that the materials of filaments come from the chromosphere. Hence the helium abundance is higher inside filaments. In this scenario, the helium abundance of cold-SFRs should be also higher if the cold-SFRs are connected with small filament or prominence eruptions. However, we have no conclusive evidence on the material sources of filaments \citep{2017ApJ...836L..11S}. The materials of filament may come from the chromosphere with high A$_{He}$ \citep{1998ApJ...494..450S} and/or originate from corona with low A$_{He}$ \citep{1993AdSpR..13i..95D}. Hence, the lower A$_{He}$ inside cold-SFRs does not mean that the cold-SFRs are not connected with filament eruptions.
{The fact that the cold-SFRs decrease with duration can be explained reasonably by the notion that cold-SFRs are associated with small cold filament or prominence eruptions. As the previous statistical analysis showed that the occurrence rate of small filament eruptions is inversely related to the scale of the eruptions \citep{2016ApJ...828L...9S,2020ApJ...889..187S}.
Magnetic erosion can also contribute to the generation of cold-SFRs. If the ICMEs contain significant cold-filament or prominence materials, the generated SFRs should be cold-SFRs associated with lower charge states and FIP bias. But the magnetic erosion process seems not to be a major contribution since cold-SFRs decrease with duration as shown in Figure~\ref{fig:2}(b).} 
%The proportion of cold-SFRs should increase with duration if the cold-SFRs are mainly generated from ICMEs by magnetic erosion process.}

The characteristics of normal-SFRs can be reasonably interpreted by the two originations, from the Sun and generated in the heliosphere both. 
If the normal-SFRs originate from the Sun and they propagate to the heliosphere with the ambient solar, we can derive the same compositional signatures between the normal-SFRs and background solar wind. On the other hand, the characteristics of normal-SFRs in different types of solar wind can also be explained reasonably by the notion that the normal-SFRs are generated in the heliosphere. In this scenario, the flux ropes are generated locally and the charge states and element abundances are not changed during the formation process of SFRs. Hence, the charge states and element abundances of normal-SFRs immersed in CH, AR, and QS solar wind are significantly different, but they are basically the same as those of their associated background solar wind.

\section{Summary and Conclusions} \label{sec:summary}

In the present study, the SFRs immersed in CH, AR, and QS solar wind are studied. The SFRs are further categorized into hot-SFRs, cold-SFRs, and normal-SFRs, according to whether or not the O$^{7+}$/O$^{6+}$ is clearly elevated or dropped inside SFRs {as} compared with background solar wind. Then the properties of hot-SFRs, cold-SFRs, normal-SFRs, and their associated background in three types of solar wind are analyzed and compared. Finally, the distributions in the space of speed versus O$^{7+}$/O$^{6+}$ for hot-SFRs, cold-SFRs, normal-SFRs, and their associated background solar wind are investigated. The main results are concluded in the following: 

\begin{enumerate}
\item The properties of SFRs immersed in CH, AR, and QS solar wind are significantly different from each other. There are small differences between the compositional parameters inside SFRs and their associated background in all three types of solar wind. The average values of compositional signatures, such as O$^{7+}$/O$^{6+}$, A$_{He}$, and FIP bias, inside SFRs are all slightly (10$\%$) higher than those inside associated background inside all three types of solar wind.
\item The proportions of hot-SFRs (cold-SFRs) increase (decrease) with the increasing of duration. The proportion of hot-SFRs is about 15$\%$ (30-40$\%$) for the SFRs shorter (longer) than 6 hours. The percentage of cold-SFRs is nearly 20$\%$ for the SFRs shorter than 3 hours. While it decreases to about 10$\%$ for the cold-SFRs ranging from 7-9 hours. No cold-SFRs last for more than 9 hours. The proportions of hot-SFRs and cold-SFRs are both slightly higher inside CH wind.
\item The properties of hot-SFRs, cold-SFRs, and normal-SFRs are quite different. The parameters of normal-SFRs are almost the same as those of the background in all three types of solar wind. Statistically, the hot-SFRs (cold-SFRs) are associated with longer (shorter) duration, lower (higher) speeds and proton temperatures, higher (lower) charge states, helium abundance, and FIP bias {as} compared with normal-SFRs and background solar wind. The same characteristic of hot-SFRs and cold-SFRs is that the magnetic field strengths are both higher than those of normal-SFRs.
\item The anti-correlations between speed and O$^{7+}$/O$^{6+}$ inside hot-SFRs (normal-SFRs) are different from (similar to) those in background solar wind. The CCs and fitting results of normal-SFRs are nearly the same as those of background solar wind. The CCs of hot-SFRs are lower than those of normal-SFRs and background solar wind. In addition, the slopes and intercepts of hot-SFRs are also different from those of normal-SFRs and background solar wind. The CCs and slops of cold-SFRs are nearly the same as those of associated background solar wind while the intercepts are all lower than those of background in all three types of solar wind.
\end{enumerate}
The present study supplies new observational facts about the properties of SFRs. The relevant results would be useful for clarifying the origination and generation mechanisms of SFRs. 
{The results indicate that most of hot-SFRs and cold-SFRs come from the Sun. The hot-SFRs may come from the streamers associated with plasma blobs and/or be produced by small-scale activities on the Sun.} 
%The cold-SFRs \textbf{may} originate from the Sun and be associated with lower \textbf{electron} temperature in source regions. 
{The cold-SFRs may be accompanied by small-scale eruptions with lower-temperature materials in the source regions. Both hot-SFRs and cold-SFRs could also be formed by magnetic erosions of ICMEs that do not contain or contain cold-filament or prominence materials in the source regions.}
%The results indicate that the hot-SFRs and cold-SFRs should both come from the Sun. The hot-SFRs may come from the streamers associated with plasma blobs and/or be produced by small-scale activities on the Sun. The cold-SFRs should also originate from the Sun and be associated with lower plasma temperature in source regions. 
The characteristics of normal-SFRs in different types of solar wind can be reasonably interpreted by the two originations, from the Sun and generated in the heliosphere both.

\begin{acknowledgments}

The authors thank the anonymous referee very much for very helpful and constructive comments and suggestions.
We thank the ACE SWICS, MAG, and SWEPAM instrument teams and the ACE Science Center for providing the ACE data.
This research is supported by the National Natural Science Foundation of China (42230203, 41974201, U1931105).
%41604147, 41627806).
\end{acknowledgments}

\bibliography{references}{}
\bibliographystyle{aasjournal}
\end{document}